\documentclass[reprint,aps,prl,twocolumn]{revtex4-1}
\usepackage{amsmath}
\usepackage{amsfonts}
\usepackage{amssymb}
\usepackage{graphicx}
\usepackage[colorlinks,linkcolor=blue,anchorcolor=blue,citecolor=blue,urlcolor=black]{hyperref}
\usepackage{mathrsfs}
\usepackage{dcolumn}
\usepackage{bm}
\usepackage{epsfig}
\usepackage{textcomp}

\begin{document}
\title{Spontaneous $\mathcal{T}$-symmetry breaking and exceptional points in cavity quantum electrodynamics systems }
\author{Yu-Kun Lu$^{1,2}$}
\author{Pai Peng$^{1,3}$}
\author{Qi-Tao Cao$^{1,2}$}
\author{Da Xu$^{1,2}$}
\author{Jan Wiersig$^4$}
\author{Qihuang Gong$^{1,2}$}
\author{Yun-Feng Xiao$^{1,2}$}
\email{Corresponding author: yfxiao@pku.edu.cn}
\altaffiliation{URL: \url{www.phy.pku.edu.cn/~yfxiao/}}
\affiliation{$^{1}$State Key Laboratory for Mesoscopic Physics and Collaborative Innovation Center of Quantum Matter, School of Physics, Peking University, Beijing 100871, People’s Republic of China}
\affiliation{$^{2}$Collaborative Innovation Center of Extreme Optics, Shanxi University, Taiyuan 030006, People’s Republic of China}
\affiliation{$^3$Department of Electrical Engineering and Computer Science, Massachusetts Institute of Technology,
Cambridge, Massachusetts 02139, USA}
\affiliation{$^4$Institut f\"ur Physik, Otto-von-Guericke-Universit\"at Magdeburg, Postfach 4120, D-39016 Magdeburg, Germany}

\date{\today}

\begin{abstract}
Spontaneous symmetry breaking has revolutionized the understanding in numerous fields of modern physics. Here, we theoretically demonstrate the spontaneous time-reversal symmetry breaking in a cavity quantum electrodynamics system in which an atomic ensemble interacts coherently with a single resonant cavity mode. The interacting system can be effectively described by two coupled oscillators with positive and negative mass, when the two-level atoms are prepared in their excited states. The occurrence of symmetry breaking is controlled by the atomic detuning and the coupling to the cavity mode, which naturally divides the parameter space into the symmetry broken and symmetry unbroken phases. The two phases are separated by a spectral singularity, a so-called exceptional point, where the eigenstates of the Hamiltonian coalesce. When encircling the singularity in the parameter space, the quasi-adiabatic dynamics shows chiral mode switching which enables topological manipulation of quantum states.

\end{abstract}

\maketitle
Spontaneous symmetry breaking (SSB), a phenomenon where the symmetric system produce symmetry-violating states, exists ubiquitously in diverse fields of modern physics, such as particle physics \cite{PhysRev.122.345,PhysRev.124.246,PhysRevLett.13.321,PhysRevLett.13.508}, condensed matter physics \cite{altland2010condensed}, cosmology \cite{PhysRevLett.48.1220}, and optics \cite{PhysRevLett.118.033901,del2017symmetry,hamel2015spontaneous,rodriguez2017symmetry}. One of the great triumphs of SSB is to classify different phases of matter. For instance, the paramagnetic-ferromagnetic phase transition occurs by breaking the spin-rotation symmetry \cite{schumann1994paramagnetic}, the time-crystal phase is realized by breaking the temporal translation symmetry \cite{PhysRevLett.109.160401,PhysRevLett.111.250402,PhysRevA.91.033617,PhysRevLett.117.090402,PhysRevLett.118.030401}, and the superconducting phase transition emerges by breaking the more subtle gauge symmetry~\cite{greiter2005electromagnetic}. Recently, in open (non-Hermitian) systems, parity-time ($\mathcal{PT}$) symmetry breaking has also been proposed theoretically \cite{bender1998real, mostafazadeh2002pseudo} and demonstrated experimentally in optical, microwave and acoustic systems \cite{guo2009observation, ruter2010observation, bittner2012p, chang2014parity, peng2014parity, zhu2014p, shi2016accessing}. In particular, $\mathcal{PT}$ symmetry breaking gives rise to exceptional points (EPs), which are non-Hermitian degeneracies that are not only of substantial theoretical interest \cite{PhysRevLett.120.093902, makris2008beam, chong2011p, longhi2009bloch, PhysRevLett.117.110802}, but also lead to fascinating applications such as unidirectional-invisible optical devices \cite{PhysRevLett.106.213901, feng2011nonreciprocal, regensburger2012parity}, unconventional lasers \cite{peng2014loss,Peng21062016,brandstetter2014reversing,feng2014single,Hodaei975}, highly efficient phonon-lasing \cite{PhysRevLett.113.053604}, slow light \cite{jing2015optomechanically} and highly sensitive nanoparticle detection~\cite{PhysRevLett.112.203901,PhysRevA.93.033809,hodaei2017enhanced,chen2017exceptional,chen2017exceptional}. 

\begin{figure}[!hbt]	\includegraphics[width=81mm,clip]{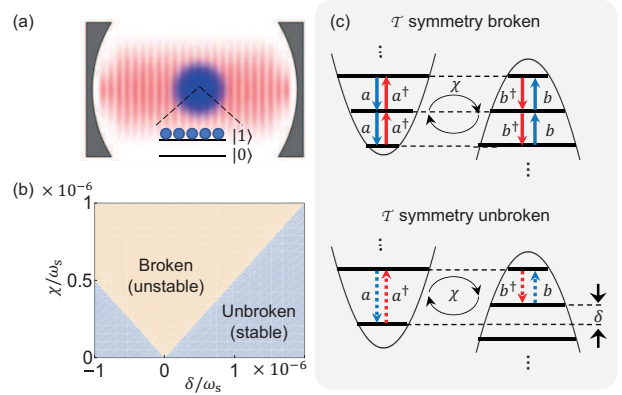}
	\caption{\label{fig1}
(color online). (a) An ensemble of two-level atoms coupled to a single-mode cavity. The atoms are initialized at their excited states.
(b) Blue (orange) region represents $\mathcal{T}$ symmetry unbroken (broken) phase with real (complex) eigenfrequencies in parameter space spanned by the effective coupling strength $\chi$ and the cavity-atom detuning $\delta$. 
(c) The system is described by two coupled oscillators with a positive mass (the cavity mode) and a negative mass (the collective spin of the atoms). The upper (bottom) panel shows $\mathcal{T}$ symmetry broken (unbroken) phase where the pair-creation term $\hat{a}^\dagger \hat{b}^\dagger$ and the pair-annihilation term $\hat{a} \hat{b}$ are on (off) resonance. The solid arrows represent the materialized processes while the dotted arrows describe virtual processes (quantum fluctuations). 
}
\end{figure}

While EPs in open systems are well understood, their existence in closed systems has been elusive. The reason is that for a closed system with an $n$-dimensional Hilbert space, the Hamiltonian has $n$ orthogonal eigenstates, which prohibit the occurrence of EPs \cite{moiseyev2011non,SI}. In this Letter, we demonstrate the spontaneous $\mathcal{T}$-symmetry breaking and the resulting EPs in a cavity quantum electrodynamics (QED) system without any gain or loss. The time-reversal operator $\mathcal{T}$ replaces $i \to - i $ while the $\mathcal{PT}$ operator replaces $i \to - i $ as well as exchanging the two modes, thus the spontaneous $\mathcal{T}$-symmetry breaking serves as the counterpart of $\mathcal{PT}$-symmetry breaking in open systems. Analogically, EPs emerge at the edge of $\mathcal{T}$-symmetry broken and unbroken phases, which is verified by the coalescence of the eigenfrequencies and the eigenmodes. In the presence of dissipations, further study reveals that the final state depends only on the chirality of the evolution trajectory encircling an EP, exhibiting the topological mode switching \cite{xu2016topological,doppler2016dynamically}. Spontaneous $\mathcal{T}$-symmetry breaking and EPs in quantum systems are of substantial interests not only for fundamental studies in physics, but also applications in various fields including quantum information processing and precise metrology.

\begin{figure}
	\includegraphics[width=85mm,clip]{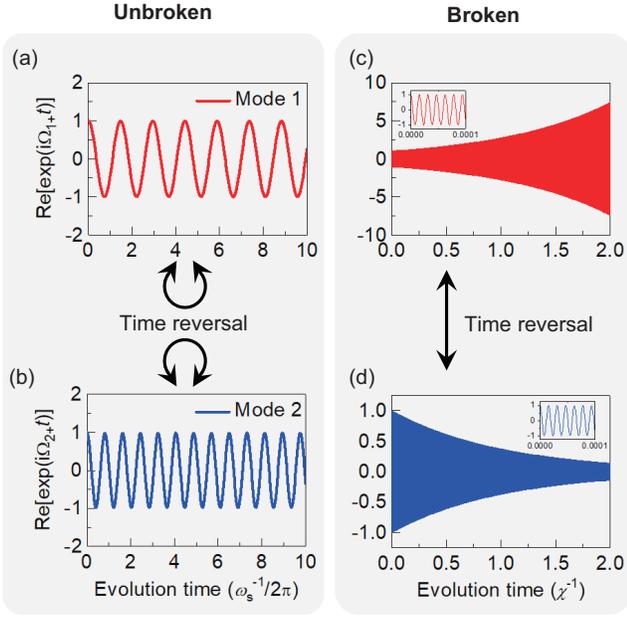}
	\caption{\label{fig3}  
	(color online).	(a), (b) Time evolution of the hybrid modes 1 and 2 in the $\mathcal{T}$-symmetry-unbroken regime, which shows that the modes are mapped onto themselves under the time-reversal operation. The parameters are chosen as $\omega_c/\omega_s=0.57$ and $\chi/\omega_s=2.5\times 10^{-4}$.
    (c), (d) Time evolution of the eigenmodes in the $\mathcal{T}$-symmetry-broken regime, which shows that the modes are mapped onto each other under the time-reversal operation. Inset is the zoomed-in view, which shows the displacement of mode 1 and mode 2 oscillates at the same frequency. The parameters are chosen as $ \omega_c/\omega_s=1$ and $\chi/\omega_s=2.5\times 10^{-4}$.
	}
\end{figure}
The system consists of $N$ identical neutral atoms interacting with a single-mode optical cavity [Fig. \ref{fig1}(a)], described by the Hamiltonian $H=\omega_c \hat{a}^\dagger \hat{a}+\omega_s \hat{S_z}+{g}(\hat{a}^\dagger+\hat{a})(\hat{S}^++\hat{S}^-)$. 
Here $\hat{a}$ ($\hat{a}^\dagger$) denotes the annihilation (creation) operator of the cavity mode, $\hat{S}_z=1/2\sum_{j=1}^{N}\sigma_z^{(j)}$ represents the collective operator of the two-level atoms with $\sigma_z^{(j)}$ being the $z$-component spin of the $j$-th atom, and $\hat{S}^+ (\hat{S}^-)$ is the collective raising (lowering) operator. 
The real parameters $\omega_c$, $\omega_s$, and $g$ represent the resonant frequency of the cavity mode, the transition frequency of the atoms, and the atom-photon coupling strength.
The atoms are assumed to be approximately in excited states $|1\rangle$ for most of the time, and their collective spin can be approximated as a harmonic oscillator with a negative mass \cite{moller2017quantum,PhysRevLett.120.013601,PhysRevLett.121.031101}, described by the bosonic operator $ \hat{b}^\dagger=\hat{S}^-/\sqrt{N}$ with a negative frequency $-\omega_s$. For a sufficiently large atom number $N$ and a weak atom-photon coupling $g$, $\hat{S}_z\approx N/2-\hat{b}^\dagger \hat{b}$ \cite{PhysRev.58.1098,RevModPhys.63.375}. The linearized Hamiltonian reads \cite{PhysRevLett.120.013601,walls2007quantum},

\begin{equation}
H=\omega_c \hat{a}^\dagger \hat{a} -\omega_{s} \hat{b}^\dagger \hat{b} + \chi(\hat{a}^\dagger +\hat{a} )(\hat{b}^\dagger +\hat{b}), 
\end{equation}
where $\chi=g\sqrt{N}$ describes the effective coupling strength.

The Heisenberg equations of the system are given by
\begin{equation}\label{eq2}
\frac{d}{dt}\begin{pmatrix}
\hat{a}\\
\hat{b} \\
\hat{a}^\dagger \\
\hat{b}^\dagger
\end{pmatrix}
=i \begin{pmatrix}
-\omega_c & -\chi    & 0    & -\chi    \\
-\chi   &  \omega_s & -\chi     & 0    \\
0   & \chi   & \omega_c  & \chi    \\
\chi   & 0    & \chi     & -\omega_s \\
\end{pmatrix}
\begin{pmatrix}
\hat{a}\\
\hat{b} \\
\hat{a}^\dagger \\
\hat{b}^\dagger
\end{pmatrix} \ .
\end{equation}
Thus, the coupled system can be described by the two hybrid eigenmodes with the eigenfrequencies satisfying
\begin{equation}\label{eq3}
\Omega_{1\pm}=\pm\frac{\sqrt{\omega_c ^2 + \omega_s ^2 +\sqrt{(\omega_c ^2 - \omega_s ^2) ^2 - 16 \chi^2 \omega_s \omega_c}} }{2},
\end{equation}
\begin{equation}\label{eq4}
\Omega_{2\pm}=\mp\frac{\sqrt{\omega_c ^2 + \omega_s ^2 -\sqrt{(\omega_c ^2 - \omega_s ^2) ^2 - 16 \chi^2 \omega_s \omega_c}}}{2},
\end{equation}
where subscripts 1 and 2 stand for the two eigenmodes, and $\Omega_{m+}$ ($\Omega_{m-}$) is the frequency of the creation (annihilation) operator of the eigenmodes.
The normalized eigenvectors corresponding to $\Omega_{m\pm}$ are denoted as $e_{m\pm}=(e_{m\pm}^1,e_{m\pm}^2,e_{m\pm}^3,e_{m\pm}^4)^T$. 

Each vector represents an operator in the basis $(\hat{a}, \hat{b},\hat{a}^\dagger, \hat{b}^\dagger)$, i.e., $\hat{e}_{m\pm}=(\hat{a}, \hat{b},\hat{a}^\dagger, \hat{b}^\dagger)\cdot e_{m\pm} $, where the $\hat{e}_{m+}$ and $\hat{e}_{m-}$ are the creation and annihilation operators of the $m$-th eigenmode satisfying $\hat e_{m+}=\hat e_{m-}^\dagger$. Specifically, the $m$-th eigenmode is the superposition of the optical and oscillator mode, with coefficients derived from $e_{m-}$ : $A_m={e_{m-}^1}{\sqrt{1-{|e_{m-}^3|^2}/{|e_{m-}^1|^2}}} $ and $B_m={e_{m-}^2}{\sqrt{1-{|e_{m-}^4|^2}/{|e_{m-}^2|^2}}}$~\cite{SI}.

\begin{figure*}
	\includegraphics[width=1.95\columnwidth,clip]{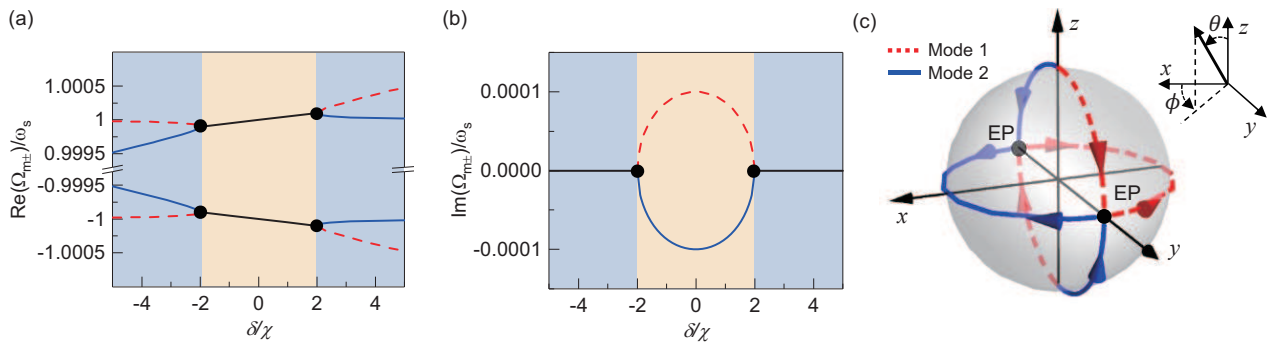}
	\caption{\label{fig2}
(color online).	(a), (b) Dependence of real and imaginary parts of the eigenfrequencies $\Omega_{m\pm}$ on the cavity-atom detuning $\delta$. Red dashed and blue solid curves stand for hybrid modes 1 and 2, respectively. The parameters are chosen as $\chi/\omega_s=10^{-4}$. 
(c) Dependence of eigenmodes on $\delta$ on a Bloch sphere, where the azimuthal angle $\phi$ denotes the relative phase, and the polar angle $\theta$ represents the relative intensity of the uncoupled cavity mode and collective spin. The black dots mark the onset of the EPs.
    }
\end{figure*}

It is clear that both the eigenfrequencies $\Omega_{m\pm}$ are real, when $(\omega_c ^2 - \omega_s ^2) ^2 - 16 \chi^2 \omega_s \omega_c>0$ corresponding to the blue region in the parameter space spanned by the coupling strength $\chi$ and the cavity-atom detuning $\delta \equiv \omega_c-\omega_s$ [see Fig. \ref{fig1}(b)]. In this case, the time evolution of the two modes exhibits harmonic oscillations, as shown in Figs. \ref{fig3}(a) and (b). The two-mode squeezing terms $\chi\hat{a}^\dagger \hat{b}^ \dagger $ and $ \chi \hat{a} \hat{b}$ merely result in quantum fluctuations (virtual processes) and cancel with each other in the sense of average [Fig. \ref{fig1}(c), bottom panel]. In this situation, the two eigenmodes can be mapped to themselves under the time-reversal operation, which preserves the $\mathcal{T}$ symmetry of the Hamiltonian \cite{SI}.

On the other hand, the eigenfrequencies become complex when $(\omega_c ^2 - \omega_s ^2) ^2 - 16 \chi^2 \omega_s \omega_c<0 $ (for example when the cavity mode is on resonance with the atoms), resulting in the instability of the system [Fig. \ref{fig1}(c), upper panel]. The instability originates from the spontaneous $\mathcal{T}$ symmetry breaking of the system.
While the Hamiltonian is invariant under the time-reversal operation, satisfying $\mathcal{T}H\mathcal{T}^{-1}=H$ ($\mathcal{T}$ is the time-reversal operator which replaces $i \to - i $), the individual eigenmodes are not necessarily $\mathcal{T}$-invariant, and the two-mode squeezing interactions $\chi\hat{a}^\dagger\hat{ b}^\dagger$ and $\chi\hat{a} \hat{b}$ play the key role in the spontaneous $\mathcal{T}$-symmetry breaking. Note that the spontaneous $\mathcal{T}$ symmetry breaking caused by the squeezing interaction differs from the parity symmetry breaking in previous works \cite{PhysRevLett.90.044101,PhysRevA.90.043817}, which is also a consequence of squeezing interaction. The key difference is that those previous models have to work in the ultra-strong coupling regime, and most importantly, the parity symmetry breaking does not lead to EPs. 
In this case, the energy of one mode grows exponentially while the other decays at the same rate [Figs. \ref{fig3}(c) and (d)]. Thus, the two eigenmodes are mapped onto each other by the time-reversal operation, and $\mathcal{T}$ symmetry is broken spontaneously. The above argument about $\mathcal{T}$ symmetry breaking is in analogy with $\mathcal{PT}$ symmetry breaking in Ref \cite{ruter2010observation}. Note that here the Hilbert space is infinite dimensional, and for unbounded operators in it, it is self-adjointness rather than Hermicity that guarantees the spectrum to be real \cite{hall2013quantum,hassani2013mathematical,simon2015quantum,gieres2000mathematical,gieres2000mathematical}. Thus it is reasonable for eigenfrequencies to acquire imaginary parts when the Hamiltonian, though remaining Hermitian, fails to be self-adjoint. In the parameter space in Fig. \ref{fig1}(b), the EP separates the $\mathcal{T}$-symmetric and the $\mathcal{T}$-symmetry-broken regions (phases), marking the onset of spontaneous symmetry breaking.

The exceptional curve corresponds to a critical detuning $\delta_c$ satisfying $(\omega_c ^2 - \omega_s ^2) ^2 - 16 \chi^2 \omega_s \omega_c=0 $, where two eigenfrequencies coalesce, i.e., $\Omega_{1-}=\Omega_{2+}=\sqrt{\omega_c ^2 + \omega_s ^2 }/{2}$. Since $\delta\ll\omega_{s(c)}$ in the optical domain, $\delta_c\approx \pm 2\chi$ [Fig. \ref{fig1}(b)].
When $|\delta| > |\delta_c|$, the real parts of the eigenfrequencies of the two modes show an attraction behavior instead of an anti-crossing, with the imaginary parts being zero [Fig. \ref{fig2}(a) and (b)]. When $|\delta| < |\delta_c|$, the gap between the real parts of the two eigenfrequencies closes while the imaginary parts bifurcate into a complex-conjugate pair. This kind of coalescence of eigenfrequencies is the typical characteristics of an EP.

To further confirm the occurrence of the EP, the dependence of the two eigenmodes on $\delta$ is presented on a Bloch sphere \cite{okamoto2013coherent,faust2013coherent,SI} in Fig.~\ref{fig2}(c). Each point $(\theta ,\phi)$ on the Bloch sphere represents a unique mode, where the polar angle $\theta$ and the azimuthal angle $\phi$ are obtained by the complex amplitude $A_m$ and $B_m$ with $\theta=2\arctan(|B_m/A_m|)$ and $\phi=\arg (A_m/B_m)$. In the largely detuned limit ($\delta\gg \chi$), all energy of the mode 1 (mode 2) resides in the cavity mode (atoms), and the system is located at the north pole (south pole) of the Bloch sphere. 
When $|\delta|$ decreases from infinity, the eigenmodes evolve toward the equator, and merge to a single mode at the critical detuning $\delta_c$ where two EPs appear. As $|\delta|$ decreases further, they part into two modes again [Fig. \ref{fig2}(c)].
The coalescence of the eigenfrequencies and modes directly verifies that EPs do exist in this closed system.
The EPs in the parameter space form an ``exceptional curve" which is exactly the critical curve in Fig.~\ref{fig1}(b). 
\begin{figure}
	\begin{center}
 \includegraphics[width=82mm,clip]{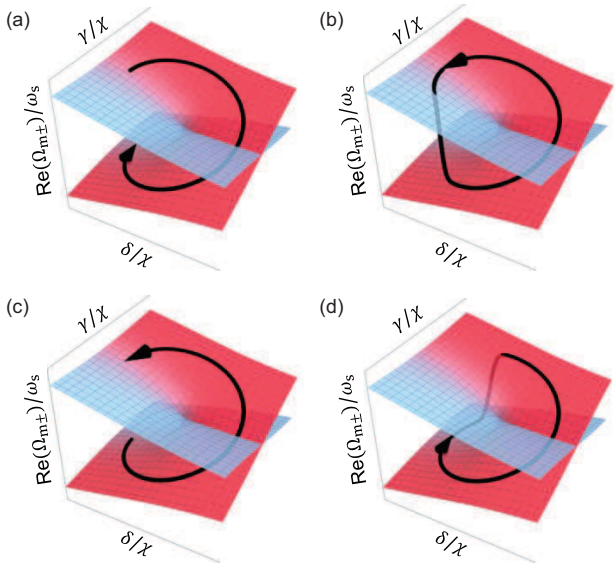}
	\caption{\label{fig4}
(color online).	Real parts of eigenfrequencies $\Omega_{m\pm}$ in the parameter space spanned by the cavity-atom detuning $\delta$ and the decay rate $\gamma$, which exhibit a Riemann surface structure. Arrows represent the evolution trajectories of the states. 
       (a) The state starts with the upper branch, clockwise.
       (b) Starts with the upper branch, counterclockwise.
       (c) Starts with the lower branch, counterclockwise.
       (d) Starts with the lower branch, clockwise. Through (a) to (d), the parameters are chosen as $\chi/\omega_s=2\times10^{-4}$, $|T|=10\chi^{-1}$, $\rho=1.5\chi$ and $\delta_0=2\chi$.
       	}
  \end{center}
\end{figure}
In a realistic system, the inescapable coupling to the environment leads to dissipation of the cavity mode and atoms with decay rate $\kappa$ and $\Gamma$, respectively. The difference between the decay rates $\gamma= \kappa-\Gamma$ provides a new degree of freedom to study the dynamics around EPs~\cite{xu2016topological,doppler2016dynamically}. As a result, the evolution matrix in Eq. (\ref{eq2}) is modified to
\begin{equation}
\label{eq5}
M= \begin{pmatrix}
-\omega_c+i\kappa & -\chi    & 0    & -\chi    \\
-\chi   &  \omega_s+i\Gamma & -\chi     & 0    \\
0   & \chi   & \omega_c+i\kappa  & \chi    \\
\chi   & 0    & \chi     & -\omega_s+i\Gamma \\
\end{pmatrix}.
\end{equation}
The real parts of the eigenfrequencies exhibit a square-root Riemann surface structure (Fig. \ref{fig4}) in the parameter space spanned by $\delta$ and $\gamma$. The evolution trajectory in the parameter space is set as a circle, i.e., $\delta(t)=\delta_0 +\rho \cos(2\pi t/T)$ and $\gamma (t)=\rho \sin (2\pi t/T)$, where $\rho$ is the radius of the circle and $T$ denotes the period of the evolution. The point ($\delta_0$, 0) is the center of the circle, which is set to the EP unless specifically mentioned. 
When the system starts with the upper (lower) branch evolving along the clockwise (counterclockwise) direction, the state remains on the Riemann surface for a sufficiently large $T$, in accordance with the adiabatic theorem [Figs. \ref{fig4}(a) and (c)].
On the other hand, when the system starts with the upper (lower) branch evolving along the trajectory counterclockwise (clockwise), the adiabatic theorem breaks down, causing the detachment from the Riemann surface even for a large $T$ [Figs. \ref{fig4}(b) and (d)].
As a result, the system always evolves to the lower (upper) branch when going clockwise (counterclockwise). This behavior can be explained by the significant amplification of one mode relative to the other mode \cite{PhysRevLett.118.093002}. For the amplified mode, its dynamical phase has a positive imaginary part which leads to the dominance in the final state. 

\begin{figure}	\includegraphics[width=80mm,clip]{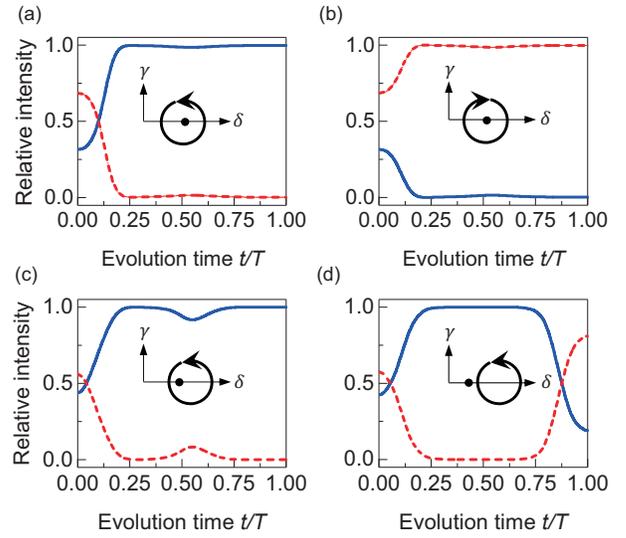}
	\caption{\label{fig5}
	(color online).	Time evolution of relative intensities of the two modes. The intensities are normalized with respect to the total intensity in two modes at each moment. 
The black point represents the EP, the arrow denotes the evolution trajectory and $T$ is the period of the evolution. The inset shows the evolution trajectory in the parameter space spanned by detuning $\delta$ and dissipation rate $\gamma$. The parameters are set as $\chi/\omega_s=2\times10^{-4}$, $\rho=1.5\chi$ and $|T|=10\chi^{-1}$.
(a) A loop centered at EP ($\delta_0=2\chi$), counterclockwise ($T=10\chi^{-1}$).
(b) A loop centered at EP ($\delta_0=2\chi$), clockwise ( $T=-10\chi^{-1}$).
(c) An excentric loop enclosing EP ($\delta_0=3.25\chi$), counterclockwise ($T=10\chi^{-1}$).
(d) A loop excluding EP ($\delta_0=6\chi$), counterclockwise ($T=10\chi^{-1}$).
}
\end{figure}

Based on the evolution in Fig. \ref{fig4}, the intensities of two instantaneous eigenmodes (see \cite{SI} for details) are quantitatively studied by normalizing with respect to the total intensity at each moment, as shown in Fig. \ref{fig5}. 
For loops centered at the EP, one mode (blue solid curve) dominates the output when going counterclockwise, while the other mode (red dashed curve) dominates the output when going clockwise [Figs. \ref{fig5}(a) and (b)].
When the loop encloses the EP excentrically, the above phenomenon remains the same as the centered case [Figs. \ref{fig5}(a) and(c)]. However, if the loop excludes the EP, the final state is no longer dominated by one state, and mode switching does not occur [Fig. \ref{fig5}(d)]. 
Thus the asymmetric state transfer is protected by the topology of the EP, immune to the smooth deformation of the evolution trajectory.

The existence of spontaneous $\mathcal{T}$-symmetry breaking holds potential for high-precision sensing. The exponentially growing mode in the unstable phase can be used as a probe: if the system is prepared in the stable phase near the exceptional curve, a weak perturbation such as the attachment of a nanoparticle can drive the system across the boundary, resulting in fast amplification of both the optical field and the oscillator motion, which has been observed experimentally \cite{PhysRevLett.120.013601,PhysRevLett.118.063604}. Note here the significant amplification originates from spontaneous $\mathcal{T}$-symmetry breaking instead of artificial gain media, so that our scheme is particularly beneficial for systems where gain is not available. Additionally, by coupling more than
two bosonic modes together, higher-order EPs can be realized, which further boosts the sensitivity~\cite{hodaei2017enhanced,SI}.

In summary, we have demonstrated the existence of spontaneous $\mathcal{T}$-symmetry breaking in closed systems without constructing the balance of gain or loss, an analogy of $\mathcal{PT}$-symmetry breaking in open systems. By showing the coalescence of eigenfrequencies as well as eigenmodes in closed cavity QED systems, it has been proved that EP emerges as a consequence of spontaneous $\mathcal{T}$-symmetry breaking. 
Furthermore, the topological nature of EP is explored, and robust mode switching is achieved by encircling the EP. 
Similar Hamiltonians can also be realized beyond cavity QED systems \cite{xiao2016optical}, such as optomechanical systems \cite{RevModPhys.86.1391,kippenberg2017arXiv,li2016quantum}, spin systems \cite{RevModPhys.63.1}, atomic systems \cite{PhysRevLett.120.013601}, Josephson junctions \cite{PhysRevLett.104.023601}, etc.
Spontaneous $\mathcal{T}$-symmetry breaking in closed systems not only broadens the understanding of SSB and singularities in quantum physics, but also reveals the rich physics in infinite dimensional systems.
Apart from its fundamental interest, spontaneous $\mathcal{T}$-symmetry breaking in closed system without gain or loss also provides a new platform for various applications, such as sensing and quantum information processing. 

The authors would like to thank Y.-C. Liu, H. Jing, Y.-X. Wang, L. Yang, S. Rotter, K. An, and H. Schomerus for fruitful discussions.
This project is supported by the National Key R$\&$D Program of China (Grant No. 2016YFA0301302), the National Natural Science Foundation of China (Grants No. 61435001, 11654003, 11474011, and 11527901), and High-performance Computing Platform of Peking University.
Y.-K. L., P. P. and Q.-T. C. contributed equally to this work. All authors contributed to the discussion and wrote the manuscript.

\bibliography{Bib}

\begin{thebibliography}{71}%
\makeatletter
\providecommand \@ifxundefined [1]{%
 \@ifx{#1\undefined}
}%
\providecommand \@ifnum [1]{%
 \ifnum #1\expandafter \@firstoftwo
 \else \expandafter \@secondoftwo
 \fi
}%
\providecommand \@ifx [1]{%
 \ifx #1\expandafter \@firstoftwo
 \else \expandafter \@secondoftwo
 \fi
}%
\providecommand \natexlab [1]{#1}%
\providecommand \enquote  [1]{``#1''}%
\providecommand \bibnamefont  [1]{#1}%
\providecommand \bibfnamefont [1]{#1}%
\providecommand \citenamefont [1]{#1}%
\providecommand \href@noop [0]{\@secondoftwo}%
\providecommand \href [0]{\begingroup \@sanitize@url \@href}%
\providecommand \@href[1]{\@@startlink{#1}\@@href}%
\providecommand \@@href[1]{\endgroup#1\@@endlink}%
\providecommand \@sanitize@url [0]{\catcode `\\12\catcode `\$12\catcode
  `\&12\catcode `\#12\catcode `\^12\catcode `\_12\catcode `\%12\relax}%
\providecommand \@@startlink[1]{}%
\providecommand \@@endlink[0]{}%
\providecommand \url  [0]{\begingroup\@sanitize@url \@url }%
\providecommand \@url [1]{\endgroup\@href {#1}{\urlprefix }}%
\providecommand \urlprefix  [0]{URL }%
\providecommand \Eprint [0]{\href }%
\providecommand \doibase [0]{http://dx.doi.org/}%
\providecommand \selectlanguage [0]{\@gobble}%
\providecommand \bibinfo  [0]{\@secondoftwo}%
\providecommand \bibfield  [0]{\@secondoftwo}%
\providecommand \translation [1]{[#1]}%
\providecommand \BibitemOpen [0]{}%
\providecommand \bibitemStop [0]{}%
\providecommand \bibitemNoStop [0]{.\EOS\space}%
\providecommand \EOS [0]{\spacefactor3000\relax}%
\providecommand \BibitemShut  [1]{\csname bibitem#1\endcsname}%
\let\auto@bib@innerbib\@empty
\bibitem [{\citenamefont {Nambu}\ and\ \citenamefont
  {Jona-Lasinio}(1961{\natexlab{a}})}]{PhysRev.122.345}%
  \BibitemOpen
  \bibfield  {author} {\bibinfo {author} {\bibfnamefont {Y.}~\bibnamefont
  {Nambu}}\ and\ \bibinfo {author} {\bibfnamefont {G.}~\bibnamefont
  {Jona-Lasinio}},\ }\href {\doibase 10.1103/PhysRev.122.345} {\bibfield
  {journal} {\bibinfo  {journal} {Phys. Rev.}\ }\textbf {\bibinfo {volume}
  {122}},\ \bibinfo {pages} {345} (\bibinfo {year}
  {1961}{\natexlab{a}})}\BibitemShut {NoStop}%
\bibitem [{\citenamefont {Nambu}\ and\ \citenamefont
  {Jona-Lasinio}(1961{\natexlab{b}})}]{PhysRev.124.246}%
  \BibitemOpen
  \bibfield  {author} {\bibinfo {author} {\bibfnamefont {Y.}~\bibnamefont
  {Nambu}}\ and\ \bibinfo {author} {\bibfnamefont {G.}~\bibnamefont
  {Jona-Lasinio}},\ }\href {\doibase 10.1103/PhysRev.124.246} {\bibfield
  {journal} {\bibinfo  {journal} {Phys. Rev.}\ }\textbf {\bibinfo {volume}
  {124}},\ \bibinfo {pages} {246} (\bibinfo {year}
  {1961}{\natexlab{b}})}\BibitemShut {NoStop}%
\bibitem [{\citenamefont {Englert}\ and\ \citenamefont
  {Brout}(1964)}]{PhysRevLett.13.321}%
  \BibitemOpen
  \bibfield  {author} {\bibinfo {author} {\bibfnamefont {F.}~\bibnamefont
  {Englert}}\ and\ \bibinfo {author} {\bibfnamefont {R.}~\bibnamefont
  {Brout}},\ }\href {\doibase 10.1103/PhysRevLett.13.321} {\bibfield  {journal}
  {\bibinfo  {journal} {Phys. Rev. Lett.}\ }\textbf {\bibinfo {volume} {13}},\
  \bibinfo {pages} {321} (\bibinfo {year} {1964})}\BibitemShut {NoStop}%
\bibitem [{\citenamefont {Higgs}(1964)}]{PhysRevLett.13.508}%
  \BibitemOpen
  \bibfield  {author} {\bibinfo {author} {\bibfnamefont {P.~W.}\ \bibnamefont
  {Higgs}},\ }\href {\doibase 10.1103/PhysRevLett.13.508} {\bibfield  {journal}
  {\bibinfo  {journal} {Phys. Rev. Lett.}\ }\textbf {\bibinfo {volume} {13}},\
  \bibinfo {pages} {508} (\bibinfo {year} {1964})}\BibitemShut {NoStop}%
\bibitem [{\citenamefont {Altland}\ and\ \citenamefont
  {Simons}(2010)}]{altland2010condensed}%
  \BibitemOpen
  \bibfield  {author} {\bibinfo {author} {\bibfnamefont {A.}~\bibnamefont
  {Altland}}\ and\ \bibinfo {author} {\bibfnamefont {B.~D.}\ \bibnamefont
  {Simons}},\ }\href@noop {} {\emph {\bibinfo {title} {Condensed matter field
  theory}}}\ (\bibinfo  {publisher} {Cambridge University Press},\ \bibinfo
  {year} {2010})\BibitemShut {NoStop}%
\bibitem [{\citenamefont {Albrecht}\ and\ \citenamefont
  {Steinhardt}(1982)}]{PhysRevLett.48.1220}%
  \BibitemOpen
  \bibfield  {author} {\bibinfo {author} {\bibfnamefont {A.}~\bibnamefont
  {Albrecht}}\ and\ \bibinfo {author} {\bibfnamefont {P.~J.}\ \bibnamefont
  {Steinhardt}},\ }\href {\doibase 10.1103/PhysRevLett.48.1220} {\bibfield
  {journal} {\bibinfo  {journal} {Phys. Rev. Lett.}\ }\textbf {\bibinfo
  {volume} {48}},\ \bibinfo {pages} {1220} (\bibinfo {year}
  {1982})}\BibitemShut {NoStop}%
\bibitem [{\citenamefont {Cao}\ \emph {et~al.}(2017)\citenamefont {Cao},
  \citenamefont {Wang}, \citenamefont {Dong}, \citenamefont {Jing},
  \citenamefont {Liu}, \citenamefont {Chen}, \citenamefont {Ge}, \citenamefont
  {Gong},\ and\ \citenamefont {Xiao}}]{PhysRevLett.118.033901}%
  \BibitemOpen
  \bibfield  {author} {\bibinfo {author} {\bibfnamefont {Q.-T.}\ \bibnamefont
  {Cao}}, \bibinfo {author} {\bibfnamefont {H.}~\bibnamefont {Wang}}, \bibinfo
  {author} {\bibfnamefont {C.-H.}\ \bibnamefont {Dong}}, \bibinfo {author}
  {\bibfnamefont {H.}~\bibnamefont {Jing}}, \bibinfo {author} {\bibfnamefont
  {R.-S.}\ \bibnamefont {Liu}}, \bibinfo {author} {\bibfnamefont
  {X.}~\bibnamefont {Chen}}, \bibinfo {author} {\bibfnamefont {L.}~\bibnamefont
  {Ge}}, \bibinfo {author} {\bibfnamefont {Q.}~\bibnamefont {Gong}}, \ and\
  \bibinfo {author} {\bibfnamefont {Y.-F.}\ \bibnamefont {Xiao}},\ }\href
  {\doibase 10.1103/PhysRevLett.118.033901} {\bibfield  {journal} {\bibinfo
  {journal} {Phys. Rev. Lett.}\ }\textbf {\bibinfo {volume} {118}},\ \bibinfo
  {pages} {033901} (\bibinfo {year} {2017})}\BibitemShut {NoStop}%
\bibitem [{\citenamefont {Del~Bino}\ \emph {et~al.}(2017)\citenamefont
  {Del~Bino}, \citenamefont {Silver}, \citenamefont {Stebbings},\ and\
  \citenamefont {Del'Haye}}]{del2017symmetry}%
  \BibitemOpen
  \bibfield  {author} {\bibinfo {author} {\bibfnamefont {L.}~\bibnamefont
  {Del~Bino}}, \bibinfo {author} {\bibfnamefont {J.~M.}\ \bibnamefont
  {Silver}}, \bibinfo {author} {\bibfnamefont {S.~L.}\ \bibnamefont
  {Stebbings}}, \ and\ \bibinfo {author} {\bibfnamefont {P.}~\bibnamefont
  {Del'Haye}},\ }\href@noop {} {\bibfield  {journal} {\bibinfo  {journal}
  {Scientific Reports}\ }\textbf {\bibinfo {volume} {7}},\ \bibinfo {pages}
  {43142} (\bibinfo {year} {2017})}\BibitemShut {NoStop}%
\bibitem [{\citenamefont {Hamel}\ \emph {et~al.}(2015)\citenamefont {Hamel},
  \citenamefont {Haddadi}, \citenamefont {Raineri}, \citenamefont {Monnier},
  \citenamefont {Beaudoin}, \citenamefont {Sagnes}, \citenamefont {Levenson},\
  and\ \citenamefont {Yacomotti}}]{hamel2015spontaneous}%
  \BibitemOpen
  \bibfield  {author} {\bibinfo {author} {\bibfnamefont {P.}~\bibnamefont
  {Hamel}}, \bibinfo {author} {\bibfnamefont {S.}~\bibnamefont {Haddadi}},
  \bibinfo {author} {\bibfnamefont {F.}~\bibnamefont {Raineri}}, \bibinfo
  {author} {\bibfnamefont {P.}~\bibnamefont {Monnier}}, \bibinfo {author}
  {\bibfnamefont {G.}~\bibnamefont {Beaudoin}}, \bibinfo {author}
  {\bibfnamefont {I.}~\bibnamefont {Sagnes}}, \bibinfo {author} {\bibfnamefont
  {A.}~\bibnamefont {Levenson}}, \ and\ \bibinfo {author} {\bibfnamefont
  {A.~M.}\ \bibnamefont {Yacomotti}},\ }\href@noop {} {\bibfield  {journal}
  {\bibinfo  {journal} {Nature Photonics}\ }\textbf {\bibinfo {volume} {9}},\
  \bibinfo {pages} {311} (\bibinfo {year} {2015})}\BibitemShut {NoStop}%
\bibitem [{\citenamefont {Rodr{\'\i}guez-Lara}\ \emph
  {et~al.}(2017)\citenamefont {Rodr{\'\i}guez-Lara}, \citenamefont
  {El-Ganainy},\ and\ \citenamefont {Guerrero}}]{rodriguez2017symmetry}%
  \BibitemOpen
  \bibfield  {author} {\bibinfo {author} {\bibfnamefont {B.}~\bibnamefont
  {Rodr{\'\i}guez-Lara}}, \bibinfo {author} {\bibfnamefont {R.}~\bibnamefont
  {El-Ganainy}}, \ and\ \bibinfo {author} {\bibfnamefont {J.}~\bibnamefont
  {Guerrero}},\ }\href@noop {} {\bibfield  {journal} {\bibinfo  {journal}
  {Science Bulletin}\ } (\bibinfo {year} {2017})}\BibitemShut {NoStop}%
\bibitem [{\citenamefont {Schumann}\ \emph {et~al.}(1994)\citenamefont
  {Schumann}, \citenamefont {Buckley},\ and\ \citenamefont
  {Bland}}]{schumann1994paramagnetic}%
  \BibitemOpen
  \bibfield  {author} {\bibinfo {author} {\bibfnamefont {F.}~\bibnamefont
  {Schumann}}, \bibinfo {author} {\bibfnamefont {M.}~\bibnamefont {Buckley}}, \
  and\ \bibinfo {author} {\bibfnamefont {J.}~\bibnamefont {Bland}},\
  }\href@noop {} {\bibfield  {journal} {\bibinfo  {journal} {Physical Review
  B}\ }\textbf {\bibinfo {volume} {50}},\ \bibinfo {pages} {16424} (\bibinfo
  {year} {1994})}\BibitemShut {NoStop}%
\bibitem [{\citenamefont {Wilczek}(2012)}]{PhysRevLett.109.160401}%
  \BibitemOpen
  \bibfield  {author} {\bibinfo {author} {\bibfnamefont {F.}~\bibnamefont
  {Wilczek}},\ }\href {\doibase 10.1103/PhysRevLett.109.160401} {\bibfield
  {journal} {\bibinfo  {journal} {Phys. Rev. Lett.}\ }\textbf {\bibinfo
  {volume} {109}},\ \bibinfo {pages} {160401} (\bibinfo {year}
  {2012})}\BibitemShut {NoStop}%
\bibitem [{\citenamefont {Wilczek}(2013)}]{PhysRevLett.111.250402}%
  \BibitemOpen
  \bibfield  {author} {\bibinfo {author} {\bibfnamefont {F.}~\bibnamefont
  {Wilczek}},\ }\href {\doibase 10.1103/PhysRevLett.111.250402} {\bibfield
  {journal} {\bibinfo  {journal} {Phys. Rev. Lett.}\ }\textbf {\bibinfo
  {volume} {111}},\ \bibinfo {pages} {250402} (\bibinfo {year}
  {2013})}\BibitemShut {NoStop}%
\bibitem [{\citenamefont {Sacha}(2015)}]{PhysRevA.91.033617}%
  \BibitemOpen
  \bibfield  {author} {\bibinfo {author} {\bibfnamefont {K.}~\bibnamefont
  {Sacha}},\ }\href {\doibase 10.1103/PhysRevA.91.033617} {\bibfield  {journal}
  {\bibinfo  {journal} {Phys. Rev. A}\ }\textbf {\bibinfo {volume} {91}},\
  \bibinfo {pages} {033617} (\bibinfo {year} {2015})}\BibitemShut {NoStop}%
\bibitem [{\citenamefont {Else}\ \emph {et~al.}(2016)\citenamefont {Else},
  \citenamefont {Bauer},\ and\ \citenamefont {Nayak}}]{PhysRevLett.117.090402}%
  \BibitemOpen
  \bibfield  {author} {\bibinfo {author} {\bibfnamefont {D.~V.}\ \bibnamefont
  {Else}}, \bibinfo {author} {\bibfnamefont {B.}~\bibnamefont {Bauer}}, \ and\
  \bibinfo {author} {\bibfnamefont {C.}~\bibnamefont {Nayak}},\ }\href
  {\doibase 10.1103/PhysRevLett.117.090402} {\bibfield  {journal} {\bibinfo
  {journal} {Phys. Rev. Lett.}\ }\textbf {\bibinfo {volume} {117}},\ \bibinfo
  {pages} {090402} (\bibinfo {year} {2016})}\BibitemShut {NoStop}%
\bibitem [{\citenamefont {Yao}\ \emph {et~al.}(2017)\citenamefont {Yao},
  \citenamefont {Potter}, \citenamefont {Potirniche},\ and\ \citenamefont
  {Vishwanath}}]{PhysRevLett.118.030401}%
  \BibitemOpen
  \bibfield  {author} {\bibinfo {author} {\bibfnamefont {N.~Y.}\ \bibnamefont
  {Yao}}, \bibinfo {author} {\bibfnamefont {A.~C.}\ \bibnamefont {Potter}},
  \bibinfo {author} {\bibfnamefont {I.-D.}\ \bibnamefont {Potirniche}}, \ and\
  \bibinfo {author} {\bibfnamefont {A.}~\bibnamefont {Vishwanath}},\ }\href
  {\doibase 10.1103/PhysRevLett.118.030401} {\bibfield  {journal} {\bibinfo
  {journal} {Phys. Rev. Lett.}\ }\textbf {\bibinfo {volume} {118}},\ \bibinfo
  {pages} {030401} (\bibinfo {year} {2017})}\BibitemShut {NoStop}%
\bibitem [{\citenamefont {Greiter}(2005)}]{greiter2005electromagnetic}%
  \BibitemOpen
  \bibfield  {author} {\bibinfo {author} {\bibfnamefont {M.}~\bibnamefont
  {Greiter}},\ }\href@noop {} {\bibfield  {journal} {\bibinfo  {journal}
  {Annals of Physics}\ }\textbf {\bibinfo {volume} {319}},\ \bibinfo {pages}
  {217} (\bibinfo {year} {2005})}\BibitemShut {NoStop}%
\bibitem [{\citenamefont {Bender}\ and\ \citenamefont
  {Boettcher}(1998)}]{bender1998real}%
  \BibitemOpen
  \bibfield  {author} {\bibinfo {author} {\bibfnamefont {C.~M.}\ \bibnamefont
  {Bender}}\ and\ \bibinfo {author} {\bibfnamefont {S.}~\bibnamefont
  {Boettcher}},\ }\href@noop {} {\bibfield  {journal} {\bibinfo  {journal}
  {Physical Review Letters}\ }\textbf {\bibinfo {volume} {80}},\ \bibinfo
  {pages} {5243} (\bibinfo {year} {1998})}\BibitemShut {NoStop}%
\bibitem [{\citenamefont {Mostafazadeh}(2002)}]{mostafazadeh2002pseudo}%
  \BibitemOpen
  \bibfield  {author} {\bibinfo {author} {\bibfnamefont {A.}~\bibnamefont
  {Mostafazadeh}},\ }\href@noop {} {\bibfield  {journal} {\bibinfo  {journal}
  {Journal of Mathematical Physics}\ }\textbf {\bibinfo {volume} {43}},\
  \bibinfo {pages} {205} (\bibinfo {year} {2002})}\BibitemShut {NoStop}%
\bibitem [{\citenamefont {Guo}\ \emph {et~al.}(2009)\citenamefont {Guo},
  \citenamefont {Salamo}, \citenamefont {Duchesne}, \citenamefont {Morandotti},
  \citenamefont {Volatier-Ravat}, \citenamefont {Aimez}, \citenamefont
  {Siviloglou},\ and\ \citenamefont {Christodoulides}}]{guo2009observation}%
  \BibitemOpen
  \bibfield  {author} {\bibinfo {author} {\bibfnamefont {A.}~\bibnamefont
  {Guo}}, \bibinfo {author} {\bibfnamefont {G.}~\bibnamefont {Salamo}},
  \bibinfo {author} {\bibfnamefont {D.}~\bibnamefont {Duchesne}}, \bibinfo
  {author} {\bibfnamefont {R.}~\bibnamefont {Morandotti}}, \bibinfo {author}
  {\bibfnamefont {M.}~\bibnamefont {Volatier-Ravat}}, \bibinfo {author}
  {\bibfnamefont {V.}~\bibnamefont {Aimez}}, \bibinfo {author} {\bibfnamefont
  {G.}~\bibnamefont {Siviloglou}}, \ and\ \bibinfo {author} {\bibfnamefont
  {D.}~\bibnamefont {Christodoulides}},\ }\href@noop {} {\bibfield  {journal}
  {\bibinfo  {journal} {Physical Review Letters}\ }\textbf {\bibinfo {volume}
  {103}},\ \bibinfo {pages} {093902} (\bibinfo {year} {2009})}\BibitemShut
  {NoStop}%
\bibitem [{\citenamefont {R{\"u}ter}\ \emph {et~al.}(2010)\citenamefont
  {R{\"u}ter}, \citenamefont {Makris}, \citenamefont {El-Ganainy},
  \citenamefont {Christodoulides}, \citenamefont {Segev},\ and\ \citenamefont
  {Kip}}]{ruter2010observation}%
  \BibitemOpen
  \bibfield  {author} {\bibinfo {author} {\bibfnamefont {C.~E.}\ \bibnamefont
  {R{\"u}ter}}, \bibinfo {author} {\bibfnamefont {K.~G.}\ \bibnamefont
  {Makris}}, \bibinfo {author} {\bibfnamefont {R.}~\bibnamefont {El-Ganainy}},
  \bibinfo {author} {\bibfnamefont {D.~N.}\ \bibnamefont {Christodoulides}},
  \bibinfo {author} {\bibfnamefont {M.}~\bibnamefont {Segev}}, \ and\ \bibinfo
  {author} {\bibfnamefont {D.}~\bibnamefont {Kip}},\ }\href@noop {} {\bibfield
  {journal} {\bibinfo  {journal} {Nature Physics}\ }\textbf {\bibinfo {volume}
  {6}},\ \bibinfo {pages} {192} (\bibinfo {year} {2010})}\BibitemShut {NoStop}%
\bibitem [{\citenamefont {Bittner}\ \emph {et~al.}(2012)\citenamefont
  {Bittner}, \citenamefont {Dietz}, \citenamefont {G{\"u}nther}, \citenamefont
  {Harney}, \citenamefont {Miski-Oglu}, \citenamefont {Richter},\ and\
  \citenamefont {Sch{\"a}fer}}]{bittner2012p}%
  \BibitemOpen
  \bibfield  {author} {\bibinfo {author} {\bibfnamefont {S.}~\bibnamefont
  {Bittner}}, \bibinfo {author} {\bibfnamefont {B.}~\bibnamefont {Dietz}},
  \bibinfo {author} {\bibfnamefont {U.}~\bibnamefont {G{\"u}nther}}, \bibinfo
  {author} {\bibfnamefont {H.}~\bibnamefont {Harney}}, \bibinfo {author}
  {\bibfnamefont {M.}~\bibnamefont {Miski-Oglu}}, \bibinfo {author}
  {\bibfnamefont {A.}~\bibnamefont {Richter}}, \ and\ \bibinfo {author}
  {\bibfnamefont {F.}~\bibnamefont {Sch{\"a}fer}},\ }\href@noop {} {\bibfield
  {journal} {\bibinfo  {journal} {Physical review letters}\ }\textbf {\bibinfo
  {volume} {108}},\ \bibinfo {pages} {024101} (\bibinfo {year}
  {2012})}\BibitemShut {NoStop}%
\bibitem [{\citenamefont {Chang}\ \emph {et~al.}(2014)\citenamefont {Chang},
  \citenamefont {Jiang}, \citenamefont {Hua}, \citenamefont {Yang},
  \citenamefont {Wen}, \citenamefont {Jiang}, \citenamefont {Li}, \citenamefont
  {Wang},\ and\ \citenamefont {Xiao}}]{chang2014parity}%
  \BibitemOpen
  \bibfield  {author} {\bibinfo {author} {\bibfnamefont {L.}~\bibnamefont
  {Chang}}, \bibinfo {author} {\bibfnamefont {X.}~\bibnamefont {Jiang}},
  \bibinfo {author} {\bibfnamefont {S.}~\bibnamefont {Hua}}, \bibinfo {author}
  {\bibfnamefont {C.}~\bibnamefont {Yang}}, \bibinfo {author} {\bibfnamefont
  {J.}~\bibnamefont {Wen}}, \bibinfo {author} {\bibfnamefont {L.}~\bibnamefont
  {Jiang}}, \bibinfo {author} {\bibfnamefont {G.}~\bibnamefont {Li}}, \bibinfo
  {author} {\bibfnamefont {G.}~\bibnamefont {Wang}}, \ and\ \bibinfo {author}
  {\bibfnamefont {M.}~\bibnamefont {Xiao}},\ }\href@noop {} {\bibfield
  {journal} {\bibinfo  {journal} {Nature photonics}\ }\textbf {\bibinfo
  {volume} {8}},\ \bibinfo {pages} {524} (\bibinfo {year} {2014})}\BibitemShut
  {NoStop}%
\bibitem [{\citenamefont {Peng}\ \emph
  {et~al.}(2014{\natexlab{a}})\citenamefont {Peng}, \citenamefont
  {{\"O}zdemir}, \citenamefont {Lei}, \citenamefont {Monifi}, \citenamefont
  {Gianfreda}, \citenamefont {Long}, \citenamefont {Fan}, \citenamefont {Nori},
  \citenamefont {Bender},\ and\ \citenamefont {Yang}}]{peng2014parity}%
  \BibitemOpen
  \bibfield  {author} {\bibinfo {author} {\bibfnamefont {B.}~\bibnamefont
  {Peng}}, \bibinfo {author} {\bibfnamefont {{\c{S}}.~K.}\ \bibnamefont
  {{\"O}zdemir}}, \bibinfo {author} {\bibfnamefont {F.}~\bibnamefont {Lei}},
  \bibinfo {author} {\bibfnamefont {F.}~\bibnamefont {Monifi}}, \bibinfo
  {author} {\bibfnamefont {M.}~\bibnamefont {Gianfreda}}, \bibinfo {author}
  {\bibfnamefont {G.~L.}\ \bibnamefont {Long}}, \bibinfo {author}
  {\bibfnamefont {S.}~\bibnamefont {Fan}}, \bibinfo {author} {\bibfnamefont
  {F.}~\bibnamefont {Nori}}, \bibinfo {author} {\bibfnamefont {C.~M.}\
  \bibnamefont {Bender}}, \ and\ \bibinfo {author} {\bibfnamefont
  {L.}~\bibnamefont {Yang}},\ }\href@noop {} {\bibfield  {journal} {\bibinfo
  {journal} {Nature Physics}\ }\textbf {\bibinfo {volume} {10}},\ \bibinfo
  {pages} {394} (\bibinfo {year} {2014}{\natexlab{a}})}\BibitemShut {NoStop}%
\bibitem [{\citenamefont {Zhu}\ \emph {et~al.}(2014)\citenamefont {Zhu},
  \citenamefont {Ramezani}, \citenamefont {Shi}, \citenamefont {Zhu},\ and\
  \citenamefont {Zhang}}]{zhu2014p}%
  \BibitemOpen
  \bibfield  {author} {\bibinfo {author} {\bibfnamefont {X.}~\bibnamefont
  {Zhu}}, \bibinfo {author} {\bibfnamefont {H.}~\bibnamefont {Ramezani}},
  \bibinfo {author} {\bibfnamefont {C.}~\bibnamefont {Shi}}, \bibinfo {author}
  {\bibfnamefont {J.}~\bibnamefont {Zhu}}, \ and\ \bibinfo {author}
  {\bibfnamefont {X.}~\bibnamefont {Zhang}},\ }\href@noop {} {\bibfield
  {journal} {\bibinfo  {journal} {Physical Review X}\ }\textbf {\bibinfo
  {volume} {4}},\ \bibinfo {pages} {031042} (\bibinfo {year}
  {2014})}\BibitemShut {NoStop}%
\bibitem [{\citenamefont {Shi}\ \emph {et~al.}(2016)\citenamefont {Shi},
  \citenamefont {Dubois}, \citenamefont {Chen}, \citenamefont {Cheng},
  \citenamefont {Ramezani}, \citenamefont {Wang},\ and\ \citenamefont
  {Zhang}}]{shi2016accessing}%
  \BibitemOpen
  \bibfield  {author} {\bibinfo {author} {\bibfnamefont {C.}~\bibnamefont
  {Shi}}, \bibinfo {author} {\bibfnamefont {M.}~\bibnamefont {Dubois}},
  \bibinfo {author} {\bibfnamefont {Y.}~\bibnamefont {Chen}}, \bibinfo {author}
  {\bibfnamefont {L.}~\bibnamefont {Cheng}}, \bibinfo {author} {\bibfnamefont
  {H.}~\bibnamefont {Ramezani}}, \bibinfo {author} {\bibfnamefont
  {Y.}~\bibnamefont {Wang}}, \ and\ \bibinfo {author} {\bibfnamefont
  {X.}~\bibnamefont {Zhang}},\ }\href@noop {} {\bibfield  {journal} {\bibinfo
  {journal} {Nature communications}\ }\textbf {\bibinfo {volume} {7}},\
  \bibinfo {pages} {11110} (\bibinfo {year} {2016})}\BibitemShut {NoStop}%
\bibitem [{\citenamefont {Yi}\ \emph {et~al.}(2018)\citenamefont {Yi},
  \citenamefont {Kullig},\ and\ \citenamefont
  {Wiersig}}]{PhysRevLett.120.093902}%
  \BibitemOpen
  \bibfield  {author} {\bibinfo {author} {\bibfnamefont {C.-H.}\ \bibnamefont
  {Yi}}, \bibinfo {author} {\bibfnamefont {J.}~\bibnamefont {Kullig}}, \ and\
  \bibinfo {author} {\bibfnamefont {J.}~\bibnamefont {Wiersig}},\ }\href
  {\doibase 10.1103/PhysRevLett.120.093902} {\bibfield  {journal} {\bibinfo
  {journal} {Phys. Rev. Lett.}\ }\textbf {\bibinfo {volume} {120}},\ \bibinfo
  {pages} {093902} (\bibinfo {year} {2018})}\BibitemShut {NoStop}%
\bibitem [{\citenamefont {Makris}\ \emph {et~al.}(2008)\citenamefont {Makris},
  \citenamefont {El-Ganainy}, \citenamefont {Christodoulides},\ and\
  \citenamefont {Musslimani}}]{makris2008beam}%
  \BibitemOpen
  \bibfield  {author} {\bibinfo {author} {\bibfnamefont {K.~G.}\ \bibnamefont
  {Makris}}, \bibinfo {author} {\bibfnamefont {R.}~\bibnamefont {El-Ganainy}},
  \bibinfo {author} {\bibfnamefont {D.}~\bibnamefont {Christodoulides}}, \ and\
  \bibinfo {author} {\bibfnamefont {Z.~H.}\ \bibnamefont {Musslimani}},\
  }\href@noop {} {\bibfield  {journal} {\bibinfo  {journal} {Physical Review
  Letters}\ }\textbf {\bibinfo {volume} {100}},\ \bibinfo {pages} {103904}
  (\bibinfo {year} {2008})}\BibitemShut {NoStop}%
\bibitem [{\citenamefont {Chong}\ \emph {et~al.}(2011)\citenamefont {Chong},
  \citenamefont {Ge},\ and\ \citenamefont {Stone}}]{chong2011p}%
  \BibitemOpen
  \bibfield  {author} {\bibinfo {author} {\bibfnamefont {Y.}~\bibnamefont
  {Chong}}, \bibinfo {author} {\bibfnamefont {L.}~\bibnamefont {Ge}}, \ and\
  \bibinfo {author} {\bibfnamefont {A.~D.}\ \bibnamefont {Stone}},\ }\href@noop
  {} {\bibfield  {journal} {\bibinfo  {journal} {Physical Review Letters}\
  }\textbf {\bibinfo {volume} {106}},\ \bibinfo {pages} {093902} (\bibinfo
  {year} {2011})}\BibitemShut {NoStop}%
\bibitem [{\citenamefont {Longhi}(2009)}]{longhi2009bloch}%
  \BibitemOpen
  \bibfield  {author} {\bibinfo {author} {\bibfnamefont {S.}~\bibnamefont
  {Longhi}},\ }\href@noop {} {\bibfield  {journal} {\bibinfo  {journal}
  {Physical review letters}\ }\textbf {\bibinfo {volume} {103}},\ \bibinfo
  {pages} {123601} (\bibinfo {year} {2009})}\BibitemShut {NoStop}%
\bibitem [{\citenamefont {Liu}\ \emph {et~al.}(2016)\citenamefont {Liu},
  \citenamefont {Zhang}, \citenamefont {{\"O}zdemir}, \citenamefont {Peng},
  \citenamefont {Jing}, \citenamefont {L\"u}, \citenamefont {Li}, \citenamefont
  {Yang}, \citenamefont {Nori},\ and\ \citenamefont
  {Liu}}]{PhysRevLett.117.110802}%
  \BibitemOpen
  \bibfield  {author} {\bibinfo {author} {\bibfnamefont {Z.-P.}\ \bibnamefont
  {Liu}}, \bibinfo {author} {\bibfnamefont {J.}~\bibnamefont {Zhang}}, \bibinfo
  {author} {\bibfnamefont {{\c{S}}.~K.}\ \bibnamefont {{\"O}zdemir}}, \bibinfo
  {author} {\bibfnamefont {B.}~\bibnamefont {Peng}}, \bibinfo {author}
  {\bibfnamefont {H.}~\bibnamefont {Jing}}, \bibinfo {author} {\bibfnamefont
  {X.-Y.}\ \bibnamefont {L\"u}}, \bibinfo {author} {\bibfnamefont {C.-W.}\
  \bibnamefont {Li}}, \bibinfo {author} {\bibfnamefont {L.}~\bibnamefont
  {Yang}}, \bibinfo {author} {\bibfnamefont {F.}~\bibnamefont {Nori}}, \ and\
  \bibinfo {author} {\bibfnamefont {Y.-x.}\ \bibnamefont {Liu}},\ }\href
  {\doibase 10.1103/PhysRevLett.117.110802} {\bibfield  {journal} {\bibinfo
  {journal} {Phys. Rev. Lett.}\ }\textbf {\bibinfo {volume} {117}},\ \bibinfo
  {pages} {110802} (\bibinfo {year} {2016})}\BibitemShut {NoStop}%
\bibitem [{\citenamefont {Lin}\ \emph {et~al.}(2011)\citenamefont {Lin},
  \citenamefont {Ramezani}, \citenamefont {Eichelkraut}, \citenamefont
  {Kottos}, \citenamefont {Cao},\ and\ \citenamefont
  {Christodoulides}}]{PhysRevLett.106.213901}%
  \BibitemOpen
  \bibfield  {author} {\bibinfo {author} {\bibfnamefont {Z.}~\bibnamefont
  {Lin}}, \bibinfo {author} {\bibfnamefont {H.}~\bibnamefont {Ramezani}},
  \bibinfo {author} {\bibfnamefont {T.}~\bibnamefont {Eichelkraut}}, \bibinfo
  {author} {\bibfnamefont {T.}~\bibnamefont {Kottos}}, \bibinfo {author}
  {\bibfnamefont {H.}~\bibnamefont {Cao}}, \ and\ \bibinfo {author}
  {\bibfnamefont {D.~N.}\ \bibnamefont {Christodoulides}},\ }\href {\doibase
  10.1103/PhysRevLett.106.213901} {\bibfield  {journal} {\bibinfo  {journal}
  {Phys. Rev. Lett.}\ }\textbf {\bibinfo {volume} {106}},\ \bibinfo {pages}
  {213901} (\bibinfo {year} {2011})}\BibitemShut {NoStop}%
\bibitem [{\citenamefont {Feng}\ \emph {et~al.}(2011)\citenamefont {Feng},
  \citenamefont {Ayache}, \citenamefont {Huang}, \citenamefont {Xu},
  \citenamefont {Lu}, \citenamefont {Chen}, \citenamefont {Fainman},\ and\
  \citenamefont {Scherer}}]{feng2011nonreciprocal}%
  \BibitemOpen
  \bibfield  {author} {\bibinfo {author} {\bibfnamefont {L.}~\bibnamefont
  {Feng}}, \bibinfo {author} {\bibfnamefont {M.}~\bibnamefont {Ayache}},
  \bibinfo {author} {\bibfnamefont {J.}~\bibnamefont {Huang}}, \bibinfo
  {author} {\bibfnamefont {Y.-L.}\ \bibnamefont {Xu}}, \bibinfo {author}
  {\bibfnamefont {M.-H.}\ \bibnamefont {Lu}}, \bibinfo {author} {\bibfnamefont
  {Y.-F.}\ \bibnamefont {Chen}}, \bibinfo {author} {\bibfnamefont
  {Y.}~\bibnamefont {Fainman}}, \ and\ \bibinfo {author} {\bibfnamefont
  {A.}~\bibnamefont {Scherer}},\ }\href@noop {} {\bibfield  {journal} {\bibinfo
   {journal} {Science}\ }\textbf {\bibinfo {volume} {333}},\ \bibinfo {pages}
  {729} (\bibinfo {year} {2011})}\BibitemShut {NoStop}%
\bibitem [{\citenamefont {Regensburger}\ \emph {et~al.}(2012)\citenamefont
  {Regensburger}, \citenamefont {Bersch}, \citenamefont {Miri}, \citenamefont
  {Onishchukov}, \citenamefont {Christodoulides},\ and\ \citenamefont
  {Peschel}}]{regensburger2012parity}%
  \BibitemOpen
  \bibfield  {author} {\bibinfo {author} {\bibfnamefont {A.}~\bibnamefont
  {Regensburger}}, \bibinfo {author} {\bibfnamefont {C.}~\bibnamefont
  {Bersch}}, \bibinfo {author} {\bibfnamefont {M.-A.}\ \bibnamefont {Miri}},
  \bibinfo {author} {\bibfnamefont {G.}~\bibnamefont {Onishchukov}}, \bibinfo
  {author} {\bibfnamefont {D.~N.}\ \bibnamefont {Christodoulides}}, \ and\
  \bibinfo {author} {\bibfnamefont {U.}~\bibnamefont {Peschel}},\ }\href@noop
  {} {\bibfield  {journal} {\bibinfo  {journal} {Nature}\ }\textbf {\bibinfo
  {volume} {488}},\ \bibinfo {pages} {167} (\bibinfo {year}
  {2012})}\BibitemShut {NoStop}%
\bibitem [{\citenamefont {Peng}\ \emph
  {et~al.}(2014{\natexlab{b}})\citenamefont {Peng}, \citenamefont
  {{\"O}zdemir}, \citenamefont {Rotter}, \citenamefont {Yilmaz}, \citenamefont
  {Liertzer}, \citenamefont {Monifi}, \citenamefont {Bender}, \citenamefont
  {Nori},\ and\ \citenamefont {Yang}}]{peng2014loss}%
  \BibitemOpen
  \bibfield  {author} {\bibinfo {author} {\bibfnamefont {B.}~\bibnamefont
  {Peng}}, \bibinfo {author} {\bibfnamefont {{\c{S}}.~K.}\ \bibnamefont
  {{\"O}zdemir}}, \bibinfo {author} {\bibfnamefont {S.}~\bibnamefont {Rotter}},
  \bibinfo {author} {\bibfnamefont {H.}~\bibnamefont {Yilmaz}}, \bibinfo
  {author} {\bibfnamefont {M.}~\bibnamefont {Liertzer}}, \bibinfo {author}
  {\bibfnamefont {F.}~\bibnamefont {Monifi}}, \bibinfo {author} {\bibfnamefont
  {C.}~\bibnamefont {Bender}}, \bibinfo {author} {\bibfnamefont
  {F.}~\bibnamefont {Nori}}, \ and\ \bibinfo {author} {\bibfnamefont
  {L.}~\bibnamefont {Yang}},\ }\href@noop {} {\bibfield  {journal} {\bibinfo
  {journal} {Science}\ }\textbf {\bibinfo {volume} {346}},\ \bibinfo {pages}
  {328} (\bibinfo {year} {2014}{\natexlab{b}})}\BibitemShut {NoStop}%
\bibitem [{\citenamefont {Peng}\ \emph {et~al.}(2016)\citenamefont {Peng},
  \citenamefont {{\"O}zdemir}, \citenamefont {Liertzer}, \citenamefont {Chen},
  \citenamefont {Kramer}, \citenamefont {Yılmaz}, \citenamefont {Wiersig},
  \citenamefont {Rotter},\ and\ \citenamefont {Yang}}]{Peng21062016}%
  \BibitemOpen
  \bibfield  {author} {\bibinfo {author} {\bibfnamefont {B.}~\bibnamefont
  {Peng}}, \bibinfo {author} {\bibfnamefont {{\c{S}}.~K.}\ \bibnamefont
  {{\"O}zdemir}}, \bibinfo {author} {\bibfnamefont {M.}~\bibnamefont
  {Liertzer}}, \bibinfo {author} {\bibfnamefont {W.}~\bibnamefont {Chen}},
  \bibinfo {author} {\bibfnamefont {J.}~\bibnamefont {Kramer}}, \bibinfo
  {author} {\bibfnamefont {H.}~\bibnamefont {Yılmaz}}, \bibinfo {author}
  {\bibfnamefont {J.}~\bibnamefont {Wiersig}}, \bibinfo {author} {\bibfnamefont
  {S.}~\bibnamefont {Rotter}}, \ and\ \bibinfo {author} {\bibfnamefont
  {L.}~\bibnamefont {Yang}},\ }\href {\doibase 10.1073/pnas.1603318113}
  {\bibfield  {journal} {\bibinfo  {journal} {Proc. Natl. Acad. Sci. U.S.A.}\
  }\textbf {\bibinfo {volume} {113}},\ \bibinfo {pages} {6845} (\bibinfo {year}
  {2016})}\BibitemShut {NoStop}%
\bibitem [{\citenamefont {Brandstetter}\ \emph {et~al.}(2014)\citenamefont
  {Brandstetter}, \citenamefont {Liertzer}, \citenamefont {Deutsch},
  \citenamefont {Klang}, \citenamefont {Sch{\"o}berl}, \citenamefont
  {T{\"u}reci}, \citenamefont {Strasser}, \citenamefont {Unterrainer},\ and\
  \citenamefont {Rotter}}]{brandstetter2014reversing}%
  \BibitemOpen
  \bibfield  {author} {\bibinfo {author} {\bibfnamefont {M.}~\bibnamefont
  {Brandstetter}}, \bibinfo {author} {\bibfnamefont {M.}~\bibnamefont
  {Liertzer}}, \bibinfo {author} {\bibfnamefont {C.}~\bibnamefont {Deutsch}},
  \bibinfo {author} {\bibfnamefont {P.}~\bibnamefont {Klang}}, \bibinfo
  {author} {\bibfnamefont {J.}~\bibnamefont {Sch{\"o}berl}}, \bibinfo {author}
  {\bibfnamefont {H.}~\bibnamefont {T{\"u}reci}}, \bibinfo {author}
  {\bibfnamefont {G.}~\bibnamefont {Strasser}}, \bibinfo {author}
  {\bibfnamefont {K.}~\bibnamefont {Unterrainer}}, \ and\ \bibinfo {author}
  {\bibfnamefont {S.}~\bibnamefont {Rotter}},\ }\href@noop {} {\bibfield
  {journal} {\bibinfo  {journal} {Nat. Commun.}\ }\textbf {\bibinfo {volume}
  {5}} (\bibinfo {year} {2014})}\BibitemShut {NoStop}%
\bibitem [{\citenamefont {Feng}\ \emph {et~al.}(2014)\citenamefont {Feng},
  \citenamefont {Wong}, \citenamefont {Ma}, \citenamefont {Wang},\ and\
  \citenamefont {Zhang}}]{feng2014single}%
  \BibitemOpen
  \bibfield  {author} {\bibinfo {author} {\bibfnamefont {L.}~\bibnamefont
  {Feng}}, \bibinfo {author} {\bibfnamefont {Z.~J.}\ \bibnamefont {Wong}},
  \bibinfo {author} {\bibfnamefont {R.-M.}\ \bibnamefont {Ma}}, \bibinfo
  {author} {\bibfnamefont {Y.}~\bibnamefont {Wang}}, \ and\ \bibinfo {author}
  {\bibfnamefont {X.}~\bibnamefont {Zhang}},\ }\href@noop {} {\bibfield
  {journal} {\bibinfo  {journal} {Science}\ }\textbf {\bibinfo {volume}
  {346}},\ \bibinfo {pages} {972} (\bibinfo {year} {2014})}\BibitemShut
  {NoStop}%
\bibitem [{\citenamefont {Hodaei}\ \emph {et~al.}(2014)\citenamefont {Hodaei},
  \citenamefont {Miri}, \citenamefont {Heinrich}, \citenamefont
  {Christodoulides},\ and\ \citenamefont {Khajavikhan}}]{Hodaei975}%
  \BibitemOpen
  \bibfield  {author} {\bibinfo {author} {\bibfnamefont {H.}~\bibnamefont
  {Hodaei}}, \bibinfo {author} {\bibfnamefont {M.-A.}\ \bibnamefont {Miri}},
  \bibinfo {author} {\bibfnamefont {M.}~\bibnamefont {Heinrich}}, \bibinfo
  {author} {\bibfnamefont {D.~N.}\ \bibnamefont {Christodoulides}}, \ and\
  \bibinfo {author} {\bibfnamefont {M.}~\bibnamefont {Khajavikhan}},\ }\href
  {\doibase 10.1126/science.1258480} {\bibfield  {journal} {\bibinfo  {journal}
  {Science}\ }\textbf {\bibinfo {volume} {346}},\ \bibinfo {pages} {975}
  (\bibinfo {year} {2014})}\BibitemShut {NoStop}%
\bibitem [{\citenamefont {Jing}\ \emph {et~al.}(2014)\citenamefont {Jing},
  \citenamefont {\"Ozdemir}, \citenamefont {L\"u}, \citenamefont {Zhang},
  \citenamefont {Yang},\ and\ \citenamefont {Nori}}]{PhysRevLett.113.053604}%
  \BibitemOpen
  \bibfield  {author} {\bibinfo {author} {\bibfnamefont {H.}~\bibnamefont
  {Jing}}, \bibinfo {author} {\bibfnamefont {S.~K.}\ \bibnamefont {\"Ozdemir}},
  \bibinfo {author} {\bibfnamefont {X.-Y.}\ \bibnamefont {L\"u}}, \bibinfo
  {author} {\bibfnamefont {J.}~\bibnamefont {Zhang}}, \bibinfo {author}
  {\bibfnamefont {L.}~\bibnamefont {Yang}}, \ and\ \bibinfo {author}
  {\bibfnamefont {F.}~\bibnamefont {Nori}},\ }\href {\doibase
  10.1103/PhysRevLett.113.053604} {\bibfield  {journal} {\bibinfo  {journal}
  {Phys. Rev. Lett.}\ }\textbf {\bibinfo {volume} {113}},\ \bibinfo {pages}
  {053604} (\bibinfo {year} {2014})}\BibitemShut {NoStop}%
\bibitem [{\citenamefont {Jing}\ \emph {et~al.}(2015)\citenamefont {Jing},
  \citenamefont {{\"O}zdemir}, \citenamefont {Geng}, \citenamefont {Zhang},
  \citenamefont {L{\"u}}, \citenamefont {Peng}, \citenamefont {Yang},\ and\
  \citenamefont {Nori}}]{jing2015optomechanically}%
  \BibitemOpen
  \bibfield  {author} {\bibinfo {author} {\bibfnamefont {H.}~\bibnamefont
  {Jing}}, \bibinfo {author} {\bibfnamefont {{\c{S}}.~K.}\ \bibnamefont
  {{\"O}zdemir}}, \bibinfo {author} {\bibfnamefont {Z.}~\bibnamefont {Geng}},
  \bibinfo {author} {\bibfnamefont {J.}~\bibnamefont {Zhang}}, \bibinfo
  {author} {\bibfnamefont {X.-Y.}\ \bibnamefont {L{\"u}}}, \bibinfo {author}
  {\bibfnamefont {B.}~\bibnamefont {Peng}}, \bibinfo {author} {\bibfnamefont
  {L.}~\bibnamefont {Yang}}, \ and\ \bibinfo {author} {\bibfnamefont
  {F.}~\bibnamefont {Nori}},\ }\href@noop {} {\bibfield  {journal} {\bibinfo
  {journal} {Scientific reports}\ }\textbf {\bibinfo {volume} {5}},\ \bibinfo
  {pages} {9663} (\bibinfo {year} {2015})}\BibitemShut {NoStop}%
\bibitem [{\citenamefont {Wiersig}(2014)}]{PhysRevLett.112.203901}%
  \BibitemOpen
  \bibfield  {author} {\bibinfo {author} {\bibfnamefont {J.}~\bibnamefont
  {Wiersig}},\ }\href {\doibase 10.1103/PhysRevLett.112.203901} {\bibfield
  {journal} {\bibinfo  {journal} {Phys. Rev. Lett.}\ }\textbf {\bibinfo
  {volume} {112}},\ \bibinfo {pages} {203901} (\bibinfo {year}
  {2014})}\BibitemShut {NoStop}%
\bibitem [{\citenamefont {Wiersig}(2016)}]{PhysRevA.93.033809}%
  \BibitemOpen
  \bibfield  {author} {\bibinfo {author} {\bibfnamefont {J.}~\bibnamefont
  {Wiersig}},\ }\href {\doibase 10.1103/PhysRevA.93.033809} {\bibfield
  {journal} {\bibinfo  {journal} {Phys. Rev. A}\ }\textbf {\bibinfo {volume}
  {93}},\ \bibinfo {pages} {033809} (\bibinfo {year} {2016})}\BibitemShut
  {NoStop}%
\bibitem [{\citenamefont {Hodaei}\ \emph {et~al.}(2017)\citenamefont {Hodaei},
  \citenamefont {Hassan}, \citenamefont {Wittek}, \citenamefont
  {Garcia-Gracia}, \citenamefont {El-Ganainy}, \citenamefont
  {Christodoulides},\ and\ \citenamefont {Khajavikhan}}]{hodaei2017enhanced}%
  \BibitemOpen
  \bibfield  {author} {\bibinfo {author} {\bibfnamefont {H.}~\bibnamefont
  {Hodaei}}, \bibinfo {author} {\bibfnamefont {A.~U.}\ \bibnamefont {Hassan}},
  \bibinfo {author} {\bibfnamefont {S.}~\bibnamefont {Wittek}}, \bibinfo
  {author} {\bibfnamefont {H.}~\bibnamefont {Garcia-Gracia}}, \bibinfo {author}
  {\bibfnamefont {R.}~\bibnamefont {El-Ganainy}}, \bibinfo {author}
  {\bibfnamefont {D.~N.}\ \bibnamefont {Christodoulides}}, \ and\ \bibinfo
  {author} {\bibfnamefont {M.}~\bibnamefont {Khajavikhan}},\ }\href@noop {}
  {\bibfield  {journal} {\bibinfo  {journal} {Nature}\ }\textbf {\bibinfo
  {volume} {548}},\ \bibinfo {pages} {187} (\bibinfo {year}
  {2017})}\BibitemShut {NoStop}%
\bibitem [{\citenamefont {Chen}\ \emph {et~al.}(2017)\citenamefont {Chen},
  \citenamefont {{\"O}zdemir}, \citenamefont {Zhao}, \citenamefont {Wiersig},\
  and\ \citenamefont {Yang}}]{chen2017exceptional}%
  \BibitemOpen
  \bibfield  {author} {\bibinfo {author} {\bibfnamefont {W.}~\bibnamefont
  {Chen}}, \bibinfo {author} {\bibfnamefont {{\c{S}}.~K.}\ \bibnamefont
  {{\"O}zdemir}}, \bibinfo {author} {\bibfnamefont {G.}~\bibnamefont {Zhao}},
  \bibinfo {author} {\bibfnamefont {J.}~\bibnamefont {Wiersig}}, \ and\
  \bibinfo {author} {\bibfnamefont {L.}~\bibnamefont {Yang}},\ }\href@noop {}
  {\bibfield  {journal} {\bibinfo  {journal} {Nature}\ }\textbf {\bibinfo
  {volume} {548}},\ \bibinfo {pages} {192} (\bibinfo {year}
  {2017})}\BibitemShut {NoStop}%
\bibitem [{\citenamefont {Moiseyev}(2011)}]{moiseyev2011non}%
  \BibitemOpen
  \bibfield  {author} {\bibinfo {author} {\bibfnamefont {N.}~\bibnamefont
  {Moiseyev}},\ }\href@noop {} {\emph {\bibinfo {title} {Non-Hermitian quantum
  mechanics}}}\ (\bibinfo  {publisher} {Cambridge University Press},\ \bibinfo
  {year} {2011})\BibitemShut {NoStop}%
\bibitem [{SI()}]{SI}%
  \BibitemOpen
  \href@noop {} {}\bibinfo {note} {See Supplementary Material for
  details.}\BibitemShut {Stop}%
\bibitem [{\citenamefont {Xu}\ \emph {et~al.}(2016)\citenamefont {Xu},
  \citenamefont {Mason}, \citenamefont {Jiang},\ and\ \citenamefont
  {Harris}}]{xu2016topological}%
  \BibitemOpen
  \bibfield  {author} {\bibinfo {author} {\bibfnamefont {H.}~\bibnamefont
  {Xu}}, \bibinfo {author} {\bibfnamefont {D.}~\bibnamefont {Mason}}, \bibinfo
  {author} {\bibfnamefont {L.}~\bibnamefont {Jiang}}, \ and\ \bibinfo {author}
  {\bibfnamefont {J.}~\bibnamefont {Harris}},\ }\href@noop {} {\bibfield
  {journal} {\bibinfo  {journal} {Nature}\ }\textbf {\bibinfo {volume} {537}},\
  \bibinfo {pages} {80} (\bibinfo {year} {2016})}\BibitemShut {NoStop}%
\bibitem [{\citenamefont {Doppler}\ \emph {et~al.}(2016)\citenamefont
  {Doppler}, \citenamefont {Mailybaev}, \citenamefont {B{\"o}hm}, \citenamefont
  {Kuhl}, \citenamefont {Girschik}, \citenamefont {Libisch}, \citenamefont
  {Milburn}, \citenamefont {Rabl}, \citenamefont {Moiseyev},\ and\
  \citenamefont {Rotter}}]{doppler2016dynamically}%
  \BibitemOpen
  \bibfield  {author} {\bibinfo {author} {\bibfnamefont {J.}~\bibnamefont
  {Doppler}}, \bibinfo {author} {\bibfnamefont {A.~A.}\ \bibnamefont
  {Mailybaev}}, \bibinfo {author} {\bibfnamefont {J.}~\bibnamefont {B{\"o}hm}},
  \bibinfo {author} {\bibfnamefont {U.}~\bibnamefont {Kuhl}}, \bibinfo {author}
  {\bibfnamefont {A.}~\bibnamefont {Girschik}}, \bibinfo {author}
  {\bibfnamefont {F.}~\bibnamefont {Libisch}}, \bibinfo {author} {\bibfnamefont
  {T.~J.}\ \bibnamefont {Milburn}}, \bibinfo {author} {\bibfnamefont
  {P.}~\bibnamefont {Rabl}}, \bibinfo {author} {\bibfnamefont {N.}~\bibnamefont
  {Moiseyev}}, \ and\ \bibinfo {author} {\bibfnamefont {S.}~\bibnamefont
  {Rotter}},\ }\href@noop {} {\bibfield  {journal} {\bibinfo  {journal}
  {Nature}\ }\textbf {\bibinfo {volume} {537}},\ \bibinfo {pages} {76}
  (\bibinfo {year} {2016})}\BibitemShut {NoStop}%
\bibitem [{\citenamefont {M{\o}ller}\ \emph {et~al.}(2017)\citenamefont
  {M{\o}ller}, \citenamefont {Thomas}, \citenamefont {Vasilakis}, \citenamefont
  {Zeuthen}, \citenamefont {Tsaturyan}, \citenamefont {Balabas}, \citenamefont
  {Jensen}, \citenamefont {Schliesser}, \citenamefont {Hammerer},\ and\
  \citenamefont {Polzik}}]{moller2017quantum}%
  \BibitemOpen
  \bibfield  {author} {\bibinfo {author} {\bibfnamefont {C.~B.}\ \bibnamefont
  {M{\o}ller}}, \bibinfo {author} {\bibfnamefont {R.~A.}\ \bibnamefont
  {Thomas}}, \bibinfo {author} {\bibfnamefont {G.}~\bibnamefont {Vasilakis}},
  \bibinfo {author} {\bibfnamefont {E.}~\bibnamefont {Zeuthen}}, \bibinfo
  {author} {\bibfnamefont {Y.}~\bibnamefont {Tsaturyan}}, \bibinfo {author}
  {\bibfnamefont {M.}~\bibnamefont {Balabas}}, \bibinfo {author} {\bibfnamefont
  {K.}~\bibnamefont {Jensen}}, \bibinfo {author} {\bibfnamefont
  {A.}~\bibnamefont {Schliesser}}, \bibinfo {author} {\bibfnamefont
  {K.}~\bibnamefont {Hammerer}}, \ and\ \bibinfo {author} {\bibfnamefont
  {E.~S.}\ \bibnamefont {Polzik}},\ }\href@noop {} {\bibfield  {journal}
  {\bibinfo  {journal} {Nature}\ }\textbf {\bibinfo {volume} {547}},\ \bibinfo
  {pages} {191} (\bibinfo {year} {2017})}\BibitemShut {NoStop}%
\bibitem [{\citenamefont {Kohler}\ \emph {et~al.}(2018)\citenamefont {Kohler},
  \citenamefont {Gerber}, \citenamefont {Dowd},\ and\ \citenamefont
  {Stamper-Kurn}}]{PhysRevLett.120.013601}%
  \BibitemOpen
  \bibfield  {author} {\bibinfo {author} {\bibfnamefont {J.}~\bibnamefont
  {Kohler}}, \bibinfo {author} {\bibfnamefont {J.~A.}\ \bibnamefont {Gerber}},
  \bibinfo {author} {\bibfnamefont {E.}~\bibnamefont {Dowd}}, \ and\ \bibinfo
  {author} {\bibfnamefont {D.~M.}\ \bibnamefont {Stamper-Kurn}},\ }\href
  {\doibase 10.1103/PhysRevLett.120.013601} {\bibfield  {journal} {\bibinfo
  {journal} {Phys. Rev. Lett.}\ }\textbf {\bibinfo {volume} {120}},\ \bibinfo
  {pages} {013601} (\bibinfo {year} {2018})}\BibitemShut {NoStop}%
\bibitem [{\citenamefont {Khalili}\ and\ \citenamefont
  {Polzik}(2018)}]{PhysRevLett.121.031101}%
  \BibitemOpen
  \bibfield  {author} {\bibinfo {author} {\bibfnamefont {F.~Y.}\ \bibnamefont
  {Khalili}}\ and\ \bibinfo {author} {\bibfnamefont {E.~S.}\ \bibnamefont
  {Polzik}},\ }\href {\doibase 10.1103/PhysRevLett.121.031101} {\bibfield
  {journal} {\bibinfo  {journal} {Phys. Rev. Lett.}\ }\textbf {\bibinfo
  {volume} {121}},\ \bibinfo {pages} {031101} (\bibinfo {year}
  {2018})}\BibitemShut {NoStop}%
\bibitem [{\citenamefont {Holstein}\ and\ \citenamefont
  {Primakoff}(1940)}]{PhysRev.58.1098}%
  \BibitemOpen
  \bibfield  {author} {\bibinfo {author} {\bibfnamefont {T.}~\bibnamefont
  {Holstein}}\ and\ \bibinfo {author} {\bibfnamefont {H.}~\bibnamefont
  {Primakoff}},\ }\href {\doibase 10.1103/PhysRev.58.1098} {\bibfield
  {journal} {\bibinfo  {journal} {Phys. Rev.}\ }\textbf {\bibinfo {volume}
  {58}},\ \bibinfo {pages} {1098} (\bibinfo {year} {1940})}\BibitemShut
  {NoStop}%
\bibitem [{\citenamefont {Klein}\ and\ \citenamefont
  {Marshalek}(1991)}]{RevModPhys.63.375}%
  \BibitemOpen
  \bibfield  {author} {\bibinfo {author} {\bibfnamefont {A.}~\bibnamefont
  {Klein}}\ and\ \bibinfo {author} {\bibfnamefont {E.~R.}\ \bibnamefont
  {Marshalek}},\ }\href {\doibase 10.1103/RevModPhys.63.375} {\bibfield
  {journal} {\bibinfo  {journal} {Rev. Mod. Phys.}\ }\textbf {\bibinfo {volume}
  {63}},\ \bibinfo {pages} {375} (\bibinfo {year} {1991})}\BibitemShut
  {NoStop}%
\bibitem [{\citenamefont {Walls}\ and\ \citenamefont
  {Milburn}(2007)}]{walls2007quantum}%
  \BibitemOpen
  \bibfield  {author} {\bibinfo {author} {\bibfnamefont {D.~F.}\ \bibnamefont
  {Walls}}\ and\ \bibinfo {author} {\bibfnamefont {G.~J.}\ \bibnamefont
  {Milburn}},\ }\href@noop {} {\emph {\bibinfo {title} {Quantum optics}}}\
  (\bibinfo  {publisher} {Springer Science \& Business Media},\ \bibinfo {year}
  {2007})\BibitemShut {NoStop}%
\bibitem [{\citenamefont {Emary}\ and\ \citenamefont
  {Brandes}(2003)}]{PhysRevLett.90.044101}%
  \BibitemOpen
  \bibfield  {author} {\bibinfo {author} {\bibfnamefont {C.}~\bibnamefont
  {Emary}}\ and\ \bibinfo {author} {\bibfnamefont {T.}~\bibnamefont
  {Brandes}},\ }\href {\doibase 10.1103/PhysRevLett.90.044101} {\bibfield
  {journal} {\bibinfo  {journal} {Phys. Rev. Lett.}\ }\textbf {\bibinfo
  {volume} {90}},\ \bibinfo {pages} {044101} (\bibinfo {year}
  {2003})}\BibitemShut {NoStop}%
\bibitem [{\citenamefont {Garziano}\ \emph {et~al.}(2014)\citenamefont
  {Garziano}, \citenamefont {Stassi}, \citenamefont {Ridolfo}, \citenamefont
  {Di~Stefano},\ and\ \citenamefont {Savasta}}]{PhysRevA.90.043817}%
  \BibitemOpen
  \bibfield  {author} {\bibinfo {author} {\bibfnamefont {L.}~\bibnamefont
  {Garziano}}, \bibinfo {author} {\bibfnamefont {R.}~\bibnamefont {Stassi}},
  \bibinfo {author} {\bibfnamefont {A.}~\bibnamefont {Ridolfo}}, \bibinfo
  {author} {\bibfnamefont {O.}~\bibnamefont {Di~Stefano}}, \ and\ \bibinfo
  {author} {\bibfnamefont {S.}~\bibnamefont {Savasta}},\ }\href {\doibase
  10.1103/PhysRevA.90.043817} {\bibfield  {journal} {\bibinfo  {journal} {Phys.
  Rev. A}\ }\textbf {\bibinfo {volume} {90}},\ \bibinfo {pages} {043817}
  (\bibinfo {year} {2014})}\BibitemShut {NoStop}%
\bibitem [{\citenamefont {Hall}(2013)}]{hall2013quantum}%
  \BibitemOpen
  \bibfield  {author} {\bibinfo {author} {\bibfnamefont {B.~C.}\ \bibnamefont
  {Hall}},\ }\href@noop {} {\emph {\bibinfo {title} {Quantum theory for
  mathematicians}}},\ Vol.\ \bibinfo {volume} {267}\ (\bibinfo  {publisher}
  {Springer},\ \bibinfo {year} {2013})\BibitemShut {NoStop}%
\bibitem [{\citenamefont {Hassani}(2013)}]{hassani2013mathematical}%
  \BibitemOpen
  \bibfield  {author} {\bibinfo {author} {\bibfnamefont {S.}~\bibnamefont
  {Hassani}},\ }\href@noop {} {\emph {\bibinfo {title} {Mathematical physics: a
  modern introduction to its foundations}}}\ (\bibinfo  {publisher} {Springer
  Science \& Business Media},\ \bibinfo {year} {2013})\BibitemShut {NoStop}%
\bibitem [{\citenamefont {Simon}(2015)}]{simon2015quantum}%
  \BibitemOpen
  \bibfield  {author} {\bibinfo {author} {\bibfnamefont {B.}~\bibnamefont
  {Simon}},\ }\href@noop {} {\emph {\bibinfo {title} {Quantum mechanics for
  Hamiltonians defined as quadratic forms}}}\ (\bibinfo  {publisher} {Princeton
  University Press},\ \bibinfo {year} {2015})\BibitemShut {NoStop}%
\bibitem [{\citenamefont {Gieres}(2000)}]{gieres2000mathematical}%
  \BibitemOpen
  \bibfield  {author} {\bibinfo {author} {\bibfnamefont {F.}~\bibnamefont
  {Gieres}},\ }\href@noop {} {\bibfield  {journal} {\bibinfo  {journal}
  {Reports on Progress in Physics}\ }\textbf {\bibinfo {volume} {63}},\
  \bibinfo {pages} {1893} (\bibinfo {year} {2000})}\BibitemShut {NoStop}%
\bibitem [{\citenamefont {Okamoto}\ \emph {et~al.}(2013)\citenamefont
  {Okamoto}, \citenamefont {Gourgout}, \citenamefont {Chang}, \citenamefont
  {Onomitsu}, \citenamefont {Mahboob}, \citenamefont {Chang},\ and\
  \citenamefont {Yamaguchi}}]{okamoto2013coherent}%
  \BibitemOpen
  \bibfield  {author} {\bibinfo {author} {\bibfnamefont {H.}~\bibnamefont
  {Okamoto}}, \bibinfo {author} {\bibfnamefont {A.}~\bibnamefont {Gourgout}},
  \bibinfo {author} {\bibfnamefont {C.-Y.}\ \bibnamefont {Chang}}, \bibinfo
  {author} {\bibfnamefont {K.}~\bibnamefont {Onomitsu}}, \bibinfo {author}
  {\bibfnamefont {I.}~\bibnamefont {Mahboob}}, \bibinfo {author} {\bibfnamefont
  {E.~Y.}\ \bibnamefont {Chang}}, \ and\ \bibinfo {author} {\bibfnamefont
  {H.}~\bibnamefont {Yamaguchi}},\ }\href@noop {} {\bibfield  {journal}
  {\bibinfo  {journal} {Nature Physics}\ }\textbf {\bibinfo {volume} {9}},\
  \bibinfo {pages} {480} (\bibinfo {year} {2013})}\BibitemShut {NoStop}%
\bibitem [{\citenamefont {Faust}\ \emph {et~al.}(2013)\citenamefont {Faust},
  \citenamefont {Rieger}, \citenamefont {Seitner}, \citenamefont {Kotthaus},\
  and\ \citenamefont {Weig}}]{faust2013coherent}%
  \BibitemOpen
  \bibfield  {author} {\bibinfo {author} {\bibfnamefont {T.}~\bibnamefont
  {Faust}}, \bibinfo {author} {\bibfnamefont {J.}~\bibnamefont {Rieger}},
  \bibinfo {author} {\bibfnamefont {M.~J.}\ \bibnamefont {Seitner}}, \bibinfo
  {author} {\bibfnamefont {J.~P.}\ \bibnamefont {Kotthaus}}, \ and\ \bibinfo
  {author} {\bibfnamefont {E.~M.}\ \bibnamefont {Weig}},\ }\href@noop {}
  {\bibfield  {journal} {\bibinfo  {journal} {Nature Physics}\ }\textbf
  {\bibinfo {volume} {9}},\ \bibinfo {pages} {485} (\bibinfo {year}
  {2013})}\BibitemShut {NoStop}%
\bibitem [{\citenamefont {Hassan}\ \emph {et~al.}(2017)\citenamefont {Hassan},
  \citenamefont {Zhen}, \citenamefont {Solja\ifmmode \check{c}\else
  \v{c}\fi{}i\ifmmode~\acute{c}\else \'{c}\fi{}}, \citenamefont {Khajavikhan},\
  and\ \citenamefont {Christodoulides}}]{PhysRevLett.118.093002}%
  \BibitemOpen
  \bibfield  {author} {\bibinfo {author} {\bibfnamefont {A.~U.}\ \bibnamefont
  {Hassan}}, \bibinfo {author} {\bibfnamefont {B.}~\bibnamefont {Zhen}},
  \bibinfo {author} {\bibfnamefont {M.}~\bibnamefont {Solja\ifmmode
  \check{c}\else \v{c}\fi{}i\ifmmode~\acute{c}\else \'{c}\fi{}}}, \bibinfo
  {author} {\bibfnamefont {M.}~\bibnamefont {Khajavikhan}}, \ and\ \bibinfo
  {author} {\bibfnamefont {D.~N.}\ \bibnamefont {Christodoulides}},\ }\href
  {\doibase 10.1103/PhysRevLett.118.093002} {\bibfield  {journal} {\bibinfo
  {journal} {Phys. Rev. Lett.}\ }\textbf {\bibinfo {volume} {118}},\ \bibinfo
  {pages} {093002} (\bibinfo {year} {2017})}\BibitemShut {NoStop}%
\bibitem [{\citenamefont {Kohler}\ \emph {et~al.}(2017)\citenamefont {Kohler},
  \citenamefont {Spethmann}, \citenamefont {Schreppler},\ and\ \citenamefont
  {Stamper-Kurn}}]{PhysRevLett.118.063604}%
  \BibitemOpen
  \bibfield  {author} {\bibinfo {author} {\bibfnamefont {J.}~\bibnamefont
  {Kohler}}, \bibinfo {author} {\bibfnamefont {N.}~\bibnamefont {Spethmann}},
  \bibinfo {author} {\bibfnamefont {S.}~\bibnamefont {Schreppler}}, \ and\
  \bibinfo {author} {\bibfnamefont {D.~M.}\ \bibnamefont {Stamper-Kurn}},\
  }\href {\doibase 10.1103/PhysRevLett.118.063604} {\bibfield  {journal}
  {\bibinfo  {journal} {Phys. Rev. Lett.}\ }\textbf {\bibinfo {volume} {118}},\
  \bibinfo {pages} {063604} (\bibinfo {year} {2017})}\BibitemShut {NoStop}%
\bibitem [{\citenamefont {Xiao}\ and\ \citenamefont
  {Gong}(2016)}]{xiao2016optical}%
  \BibitemOpen
  \bibfield  {author} {\bibinfo {author} {\bibfnamefont {Y.-F.}\ \bibnamefont
  {Xiao}}\ and\ \bibinfo {author} {\bibfnamefont {Q.}~\bibnamefont {Gong}},\
  }\href@noop {} {\bibfield  {journal} {\bibinfo  {journal} {Science Bulletin}\
  }\textbf {\bibinfo {volume} {61}},\ \bibinfo {pages} {185} (\bibinfo {year}
  {2016})}\BibitemShut {NoStop}%
\bibitem [{\citenamefont {Aspelmeyer}\ \emph {et~al.}(2014)\citenamefont
  {Aspelmeyer}, \citenamefont {Kippenberg},\ and\ \citenamefont
  {Marquardt}}]{RevModPhys.86.1391}%
  \BibitemOpen
  \bibfield  {author} {\bibinfo {author} {\bibfnamefont {M.}~\bibnamefont
  {Aspelmeyer}}, \bibinfo {author} {\bibfnamefont {T.~J.}\ \bibnamefont
  {Kippenberg}}, \ and\ \bibinfo {author} {\bibfnamefont {F.}~\bibnamefont
  {Marquardt}},\ }\href {\doibase 10.1103/RevModPhys.86.1391} {\bibfield
  {journal} {\bibinfo  {journal} {Rev. Mod. Phys.}\ }\textbf {\bibinfo {volume}
  {86}},\ \bibinfo {pages} {1391} (\bibinfo {year} {2014})}\BibitemShut
  {NoStop}%
\bibitem [{\citenamefont {Bernier}\ \emph {et~al.}()\citenamefont {Bernier},
  \citenamefont {T\'{o}th}, \citenamefont {Feofanov},\ and\ \citenamefont
  {Kippenberg}}]{kippenberg2017arXiv}%
  \BibitemOpen
  \bibfield  {author} {\bibinfo {author} {\bibfnamefont {N.~R.}\ \bibnamefont
  {Bernier}}, \bibinfo {author} {\bibfnamefont {L.~D.}\ \bibnamefont
  {T\'{o}th}}, \bibinfo {author} {\bibfnamefont {A.~K.}\ \bibnamefont
  {Feofanov}}, \ and\ \bibinfo {author} {\bibfnamefont {T.~J.}\ \bibnamefont
  {Kippenberg}},\ }\href {https://arxiv.org/abs/1709.02220} {\bibinfo
  {journal} {arXiv:1709.02220}\ }\BibitemShut {NoStop}%
\bibitem [{\citenamefont {Li}\ and\ \citenamefont {Yin}(2016)}]{li2016quantum}%
  \BibitemOpen
\bibfield  {journal} {  }\bibfield  {author} {\bibinfo {author} {\bibfnamefont
  {T.}~\bibnamefont {Li}}\ and\ \bibinfo {author} {\bibfnamefont {Z.-Q.}\
  \bibnamefont {Yin}},\ }\href@noop {} {\bibfield  {journal} {\bibinfo
  {journal} {Science Bulletin}\ }\textbf {\bibinfo {volume} {61}},\ \bibinfo
  {pages} {163} (\bibinfo {year} {2016})}\BibitemShut {NoStop}%
\bibitem [{\citenamefont {Manousakis}(1991)}]{RevModPhys.63.1}%
  \BibitemOpen
  \bibfield  {author} {\bibinfo {author} {\bibfnamefont {E.}~\bibnamefont
  {Manousakis}},\ }\href {\doibase 10.1103/RevModPhys.63.1} {\bibfield
  {journal} {\bibinfo  {journal} {Rev. Mod. Phys.}\ }\textbf {\bibinfo {volume}
  {63}},\ \bibinfo {pages} {1} (\bibinfo {year} {1991})}\BibitemShut {NoStop}%
\bibitem [{\citenamefont {Nataf}\ and\ \citenamefont
  {Ciuti}(2010)}]{PhysRevLett.104.023601}%
  \BibitemOpen
  \bibfield  {author} {\bibinfo {author} {\bibfnamefont {P.}~\bibnamefont
  {Nataf}}\ and\ \bibinfo {author} {\bibfnamefont {C.}~\bibnamefont {Ciuti}},\
  }\href {\doibase 10.1103/PhysRevLett.104.023601} {\bibfield  {journal}
  {\bibinfo  {journal} {Phys. Rev. Lett.}\ }\textbf {\bibinfo {volume} {104}},\
  \bibinfo {pages} {023601} (\bibinfo {year} {2010})}\BibitemShut {NoStop}%
\end{thebibliography}%

\end{document}